# Energy conditions for Inhomogeneous EOS and its Thermodynamics analysis with the resolution on finite time future singularity problems

**Alokananda Kar[1] Shouvik Sadhukhan[2] Surajit Chattopadhay[3]**

1. 2nd year M.Sc, University of Calcutta ; Department of Physics ; West Bengal ; India
   Email : alokanandakar@gmail.com
2. 2nd year M.Sc, Indian Institute of Technology ; Kharagpur ; Department of Physics ; West Bengal ; India
   Email :shouvikphysics1996@gmail.com
3. Department of Mathematics, Amity University Major Arterial Road, Action Area II, Rajarhat New Town, Kolkata 700135, India
   Email : surajitchatto@outlook.com

## Abstract

In this paper we study two different cases of inhomogeneous EOS of the form $p_d = w(t)\rho_d + w_1 f(H,t)$ . We derive the energy density of dark fluid and dark matter component for both the cases. Further we calculate the evolution of energy density, gravitational constant and cosmological constant. We also explore the finite time singularity and thermodynamic stability conditions for the two cases of EOS. Finally, we discuss the thermodynamics of inhomogeneous EOS with the derivation of internal energy, Temperature and entropy and also show that all the stability conditions are satisfied for the two cases of EOS.

**Keyword:** Cosmology, General theory of relativity, Isotropic and Anisotropic fluids, Inflation, Fluid mechanics, Bianchi Models, Viscosity, Gravitational physics, Inhomogeneous fluid

## 1. Introduction

Dark energy has become one of the most important field of research after the discovery of cosmic acceleration. The simplest model of dark energy that is cosmological constant model leads to several problems for which several other forms of fluid (phantom, quintessence, inhomogeneous fluids etc.) are considered [91-100]. Each of these models explain some specific elements of late time acceleration. Phantom era is one of the strangest era of the universe evolution. Several attempts have been made to explain phantom cosmology in the future, finite time singularity which is an essential component of phantom cosmology. For this reason, we change the type of fluid and consider inhomogeneous EOS to solve the finite time singularity. It is also assumed that modern universe is filled with some mysterious, negative pressure fluid (dark energy). The origin of this dark energy is not yet clear. We consider two different cases of a inhomogeneous EOS parameter [91-106]. We introduce the formalism of inhomogeneous EOS to flat FRW model, and study its effect on the evolution of G, energy density and Cosmological constant and also how future singularity changes. So we can easily observe that the time variation of that two constant will have considerable effect on the evolution profile of cosmology. Therefore, if we use them in the scalar field dark energy model with inhomogeneous cosmology, we'll definitely get some new ideas towards the solution of future singularity. The time dependent Λ and G may produce different conclusion on the future singularity problems.

**From the paper [108] we get the action**

$$\mathbf{S} = \int \mathbf{d^4xL} = \int \mathbf{d^4x\{\sqrt{-g}[\frac{R}{G} + F(G) + L_m]\}}$$

**We can derive the Einstein field equation with time variable Λ.**

$$\mathbf{R_{\mu\vartheta} - \frac{1}{2}Rg_{\mu\vartheta} = 8\pi GT_{\mu\vartheta} + g_{\mu\vartheta}\Lambda(t)} \qquad \text{where } \Lambda(t) = ½\, GF(G)$$

**Using this field equation, we can derive Friedman equation with variable G and Λ.**

$$\left(\frac{\dot{a}}{a}\right)^2 = \frac{8\pi G}{3}\rho + \frac{\Lambda}{3}$$

$$\left(\frac{\ddot{a}}{a}\right) = -\frac{4\pi G}{3}(\rho + 3p) + \frac{\Lambda}{3}$$

**And from the conservation principle of energy momentum tensor we can get,**

$$8\pi\dot{G}\rho_{tot} + \dot{\Lambda} + 8\pi G[\dot{\rho}_{tot} + 3H(\rho_{tot} + p_{tot})] = 0$$

**So, we observe that the Friedmann equations don't contain any term with $\dot{G}(t)$.**

The paper is organized as follows: In section 2, we calculate the energy conservation equations for different types of energy density from flat FRW model and considering inhomogeneous EOS.

In section 3 we provide the calculation of dark fluid, dark matter energy density and cosmological constants [1-50].

Section 4 provides the details of thermodynamic energy conditions and finite time singularity problems which should be satisfied [100-102].

In section 5 and 6 we derive the ideal fluid density for case I and case II of inhomogeneous EOS respectively.

In section 7 We derive the scalar field, scalar field density, scalar field potential and gravitational constant for inhomogeneous EOS [56-80].

Section 8 and 9 we analyze the finite time singularity for two different types of Inhomogeneous EOS [90-95].

In section 10 and 11 the thermodynamics and thermodynamic stability are discussed for the two cases [103-106].

## 2. Basic Calculations towards the addition of Inhomogeneous Fluid

Here the necessary field equations will be as follows;

$$\left(\frac{\dot{a}}{a}\right)^2 = \frac{8\pi G(t)}{3}\rho_{tot} + \frac{\Lambda(t)}{3} \qquad (1)$$

$$\left(\frac{\ddot{a}}{a}\right) = -\frac{4\pi G(t)}{3}(\rho_{tot} + 3p_{tot}) + \frac{\Lambda(t)}{3} \qquad (2)$$

Where

$$\rho_{tot} = \rho_m + \rho_d + \rho_\varphi + \rho_c \;; \tag{3}$$

$$p_{tot} = p_m + p_d + p_\varphi + p_c \tag{4}$$

And;

| | |
|---|---|
| $\rho_m$ = Density of Dark matter;<br>$\rho_d$ = Density of ideal fluid;<br>$\rho_\varphi$ = Density of scalar field $= \frac{1}{2}\dot{\varphi}^2 + V(\varphi)$ ;<br>$\rho_c$ = Density of Dark fluid (Chaplygin gas ); | $p_m$ = Pressure of Dark matter;<br>$p_d$ = Pressure of ideal fluid;<br>$p_\varphi$ = Pressure of scalar field $= \frac{1}{2}\dot{\varphi}^2 - V(\varphi)$ ;<br>$p_c$ = Pressure of Dark fluid (Chaplygin gas ); |

We'll use the Dark fluid pressure as follows;

$$p_c = -\frac{B}{\rho_c} \;; \tag{5}$$

We can choose the inhomogeneous EOS of ideal fluid as following format:

$$p_d = w(t)\rho_d + w_1 f(H,t) \tag{6}$$

For this format We'll use two different cases.

**Case I:** $p_d = a_2 t^{-\alpha}\rho_d - ct^{-\beta}$ (This EOS will be substituted with $p_d = a_2(a_1+bt)^{-\alpha}\rho_d - c(a_1 + bt)^{-\beta}$ to simplify our calculation. Basically, $t$ has been substituted with $a_1 + bt$ which has helped us to solve initial and finite time future singularity problems.) (7)

**Case II:** $p_d = A\rho_d + BH^2$ (8)

Now from conservation of energy momentum tensor we get the following equations;

$$\dot{\rho}_m + 3H(\rho_m + p_m) = 0 \;; \tag{9}$$

$$\dot{\rho}_d + 3H(\rho_d + p_d) = -3H\rho_d\delta \;; \tag{10}$$

$$\dot{\rho}_\varphi + 3H(\rho_\varphi + p_\varphi) = 3H\rho_d\delta \;; \tag{11}$$

$$\dot{\rho}_c + 3H(\rho_c + p_c) = 0 \;; \tag{12}$$

And also; $8\pi\dot{G}\rho_{tot} + \dot{\Lambda} = 0$ ; without viscosity case. (13)

Here $\delta$ is the interaction constant that controls the energy transfer between scalar field and ideal fluid.

In this equation density of ideal fluid will only contribute in $\rho_{tot}$ as because the variation of $\Lambda$ is independent on scalar field. The Dark Fluid density is basically the lambda correspondence on energy density. That's why $\rho_{tot}$ will be substituted by $\rho_d$. The equation will be as follows.

$$8\pi\dot{G}\rho_d + \dot{\Lambda} = 0 \;; \tag{14}$$

## 3. Calculations for Dark fluid, Dark matter energy densities and cosmological constant and also the choice of scale factor

Now we get from field equation that the lambda contribution on density can be written as;

$$H^2 = \frac{\rho_\Lambda}{3} \approx \frac{\rho_c}{3} \ ; \tag{15}$$

and also

$$3\text{H} = \sqrt{3\rho_c} \tag{16}$$

So from the relation of Dark fluid we get (using Chaplygin gas);

$$\dot{\rho}_c + 3H\left(\rho_c - \frac{B}{\rho_c}\right) = 0 \ ; \tag{17}$$

$$\text{Or; } {\rho_c}^2 = ({\rho_{c0}}^2 - B)\left(\frac{a_0}{a}\right)^6 + B \ ; \tag{18}$$

Again, we get considering H as a parameter;

$$({\rho_c}^2 - B) = ({\rho_{c0}}^2 - B)\exp(-3Ht) = ({\rho_{c0}}^2 - B)\exp(-\sqrt{3\rho_c}\,) \tag{19}$$

Now considering 1st order approximation of $\exp(-\sqrt{3\rho_c}\,)$ ; we get the quadratic equation as follows;

$${\rho_c}^2 + \sqrt{3}({\rho_{c0}}^2 - B)t\,\rho_c - {\rho_{c0}}^2 = 0 \tag{20}$$

So, the ethical solution will be

$$\rho_c = \frac{\sqrt{[\sqrt{3}({\rho_{c0}}^2 - B)t]^2 + 4{\rho_{c0}}^2}}{2} - \frac{\sqrt{3}({\rho_{c0}}^2 - B)t}{2} \tag{21}$$

Now using this density in the above $\rho_c - a$ relation we get;

$$\text{a(t)} = a_0({\rho_{c0}}^2 - B)^{\frac{1}{6}}\left(\left(\frac{\sqrt{\left[\sqrt{3}({\rho_{c0}}^2 - B)t\right]^2 + 4{\rho_{c0}}^2}}{2} - \frac{\sqrt{3}({\rho_{c0}}^2 - B)t}{2}\right)^2 - B\right)^{-\frac{1}{6}} \tag{22}$$

or in brief we can rewrite this as follows;

$$\text{a(t)} = a_{01}\left(\left(\sqrt{b_{01}t^2 + c_{01}} - d_{01}\text{t}\right)^2 - B\right)^{-\frac{1}{6}} \tag{23}$$

now using this scale factor, we now derive the Hubble parameter as follows;

$$\text{H} = \frac{\left(\sqrt{b_{01}t^2 + c_{01}} - d_{01}\text{t}\right)\left(\frac{b_{01}t}{\sqrt{b_{01}t^2 + c_{01}}} - d_{01}\right)}{3\left(\left(\sqrt{b_{01}t^2 + c_{01}} - d_{01}\text{t}\right)^2 - B\right)} \ ; \tag{24}$$

So, we see that the above expression is almost equivalent to H = $\frac{constant}{a_1 + nt}$ which is completely similar to our earlier approximation about initial singularity free scale factor. The graphical may provide the similarity about the above conclusion. So without loss of generality we may get our earlier choice of scale factor and we can proceed our calculation with the scale factor function as follows.

$$a(t) = a_{01}(a_1 + nt)^m \tag{25}$$

we have already assumed that $\Lambda=3\alpha\left(\frac{\dot{a}}{a}\right)^2 + \beta\left(\frac{\ddot{a}}{a}\right)$; so, we now use our newly approximated scale factor to get the exact time varying function of cosmological constant. As the approximated scale factor is completely similar to earlier assumption, the time dependent function of cosmological constant will also be completely same as earlier. So the cosmological constant will be as follows [60-80].

$$\Lambda = \frac{[3\alpha m^2 + m\beta(m-1)]n^2}{(a_1+nt)^2} \quad ; \tag{26}$$

From the relation of $\rho_c - a$ we get the time varying $\rho_c$ as follows;

$$\rho_c = \sqrt{(\rho_{c0}{}^2 - B)\left(\frac{a_0}{a_{01}}\right)^6 (a_1 + nt)^{-6m} + B} \tag{27}$$

The energy density for dark matter can be taken as;

$$\rho_m = \rho_{m0} a^{-3} = \rho_{m0} a_0{}^{-3} (a_1 + nt)^{-3m} \tag{28}$$

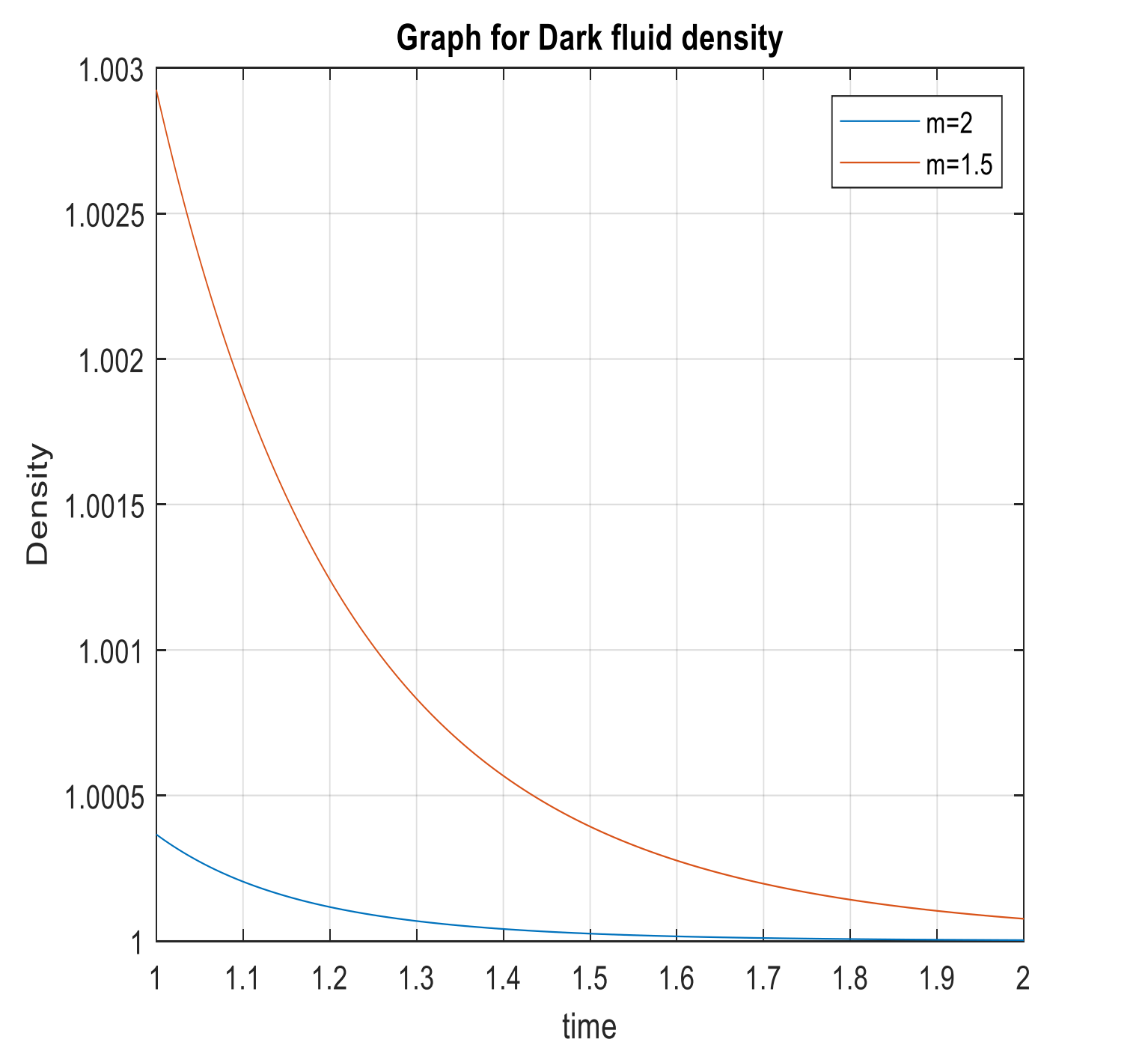

Figure 1: $\rho_c$ vs time

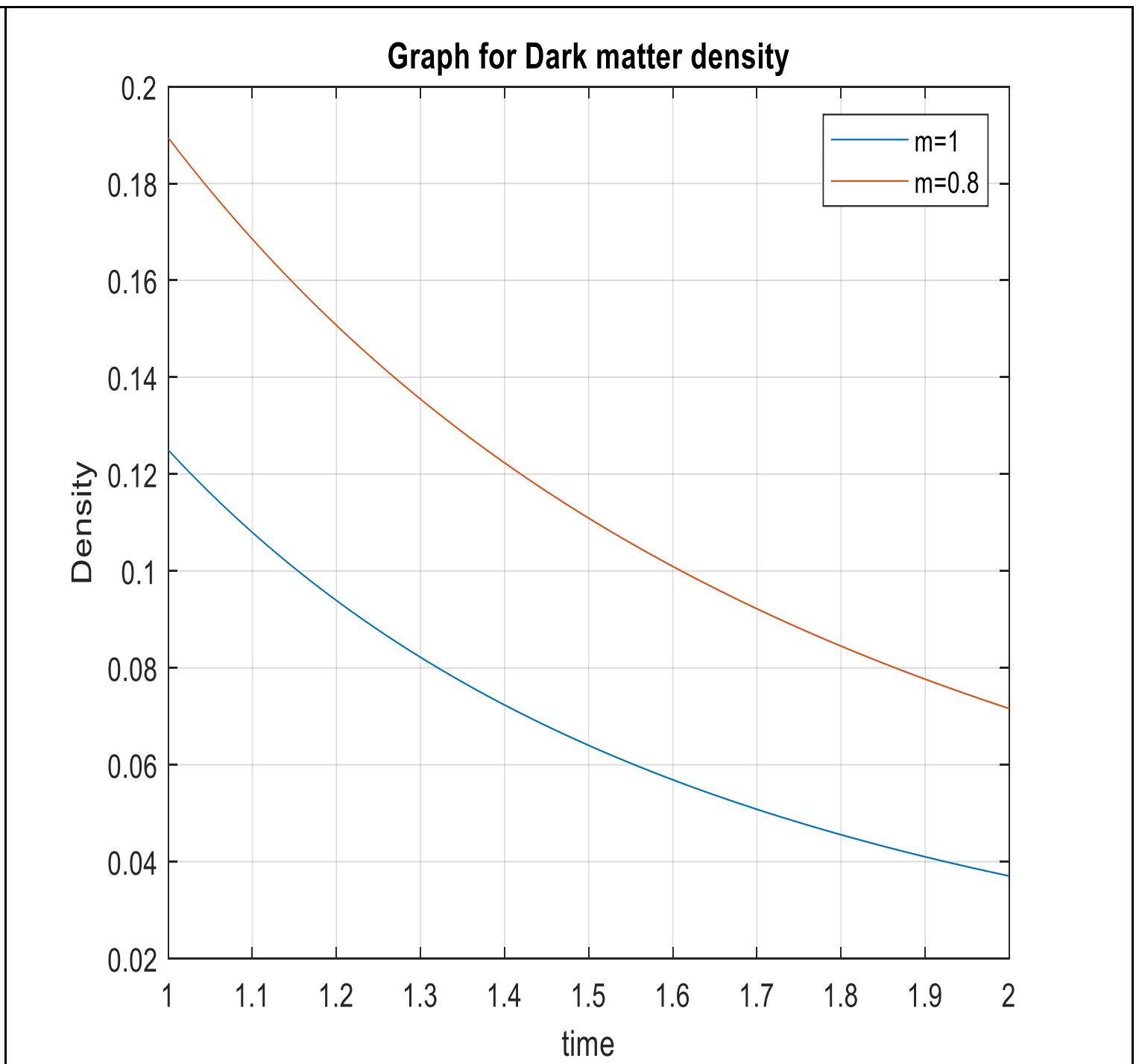

Figure 2: $\rho_m$ vs time

Here we have found that the Dark fluid density $\rho_c$ provides positive values which is changing with time. This gives the idea of dynamic cosmological constant. As we know that the cosmological constant provides negative pressure to produce accelerating expansion, this dark fluid density must provide negative pressure during cosmic expansion. Hence, the Dark fluid is used just to introduce the idea of negative pressure as well as the dynamic cosmological constant.

## 4. Basics of finite time singularity problems and the Thermodynamics Energy conditions:

For finite time singularity problems, we know the following types [94];

- Type I (known as Big Rip): At any finite time, $t = t_s$ we'll have $a \to \infty$; $\rho \to \infty$; $p \to \infty$
- Type II (known as Sudden Singularity): At any finite time, $t = t_s$ we'll have $a \to a_s$; $\rho \to \rho_s$; $p \to \infty$
- Type III: At any finite time, $t = t_s$ we'll have $a \to a_s$; $\rho \to \infty$; $p \to \infty$ ; it happens only in the EOS type of $p = - \rho - A \rho^{\alpha}$.
- Type IV: At any finite time, $t = t_s$ we'll have $a \to a_s$; $\rho \to \rho_s$; $p \to p_s$ Moreover, the Hubble rate and its first derivative also remain finite, but the higher derivatives, or some of these diverge. This kind of singularity mainly comes into play when $p = - \rho - f(\rho)$.

The thermodynamic energy conditions are basically derived from the well-known Raychaudhuri's equation. For a congruence of time-like and null-like geodesics, the Raychaudhuri equations are given in the following forms [107];

$$\frac{d\theta}{d\tau} = -\frac{1}{3}\theta^2 - \sigma_{\mu\nu}\sigma^{\mu\nu} + \omega_{\mu\nu}\omega^{\mu\nu} - R_{\mu\nu}u^{\mu}u^{\nu} \quad (29)$$

And

$$\frac{d\theta}{d\tau} = -\frac{1}{3}\theta^2 - \sigma_{\mu\nu}\sigma^{\mu\nu} + \omega_{\mu\nu}\omega^{\mu\nu} - R_{\mu\nu}n^{\mu}n^{\nu} \quad (30)$$

Where $\theta$ is the expansion factor, $n^{\mu}n^{\nu}$ is the null vector, and $\sigma_{\mu\nu}\sigma^{\mu\nu}$ and $\omega_{\mu\nu}\omega^{\mu\nu}$ are, respectively, the shear and the rotation associated with the vector field $u^{\mu}u^{\nu}$ . For attractive gravity we'll have the followings;

$$R_{\mu\nu}u^{\mu}u^{\nu} \geq 0 \;\; and \; R_{\mu\nu}n^{\mu}n^{\nu} \geq 0$$

So for our matter-fluid distribution we may write this condition as follows;

- NEC = $\rho + p \geq 0$
- WEC = $\rho \geq 0$ and $\rho + p \geq 0$
- SEC = $\rho + 3p \geq 0$ and $\rho + p \geq 0$
- DEC = $\rho \geq 0$ and $-\rho \leq p \leq \rho$

## 5. Derivations for ideal fluid energy density with Type I Inhomogeneous EOS

Now we'll derive $\rho_d$ using the inhomogeneous EOS one by one.

From our 1st type of EOS we get;

$$\dot{\rho}_d + \frac{3mn\rho_d(1+\delta)}{(a_1+nt)} + \frac{3mn\rho_d a_2}{(a_1+nt)^{1+\alpha}} - \frac{3mnc}{(a_1+nt)^{1+\beta}} = 0 \quad (31)$$

So, let's discuss about the solution of the choice of density of dark fluid from the above discussed differential equations. We get;

we consider $\alpha \neq \beta$ . So, we get;

$$\rho_d = c_1 \exp\left(-\frac{constant1*\log(a_1+nt)-\frac{constant2*(a_1+nt)^{-\alpha}}{\alpha}}{n}\right) + \frac{constant2*c}{a_2*\alpha*n} *$$
$$(\exp\left(-\frac{constant1*\log(a_1+nt)-\frac{constant2*(a_1+nt)^{-\alpha}}{\alpha}}{n}\right)) * (a_1+nt)^{-\beta+\frac{constant1}{n}} *$$
$$\left(\frac{constant2*(a_1+nt)^{-\alpha}}{\alpha*n}\right)^{\frac{-\beta+\frac{constant1}{n}}{\alpha}} Gamma(-\frac{-\beta+\frac{constant1}{n}}{\alpha}, \frac{constant2*(a_1+nt)^{-\alpha}}{\alpha*n}); \quad (32)$$

Here constant1 = $3mn(1+\delta)$; constant2 = $3mna_2$; $c_1$ = constant of integration.

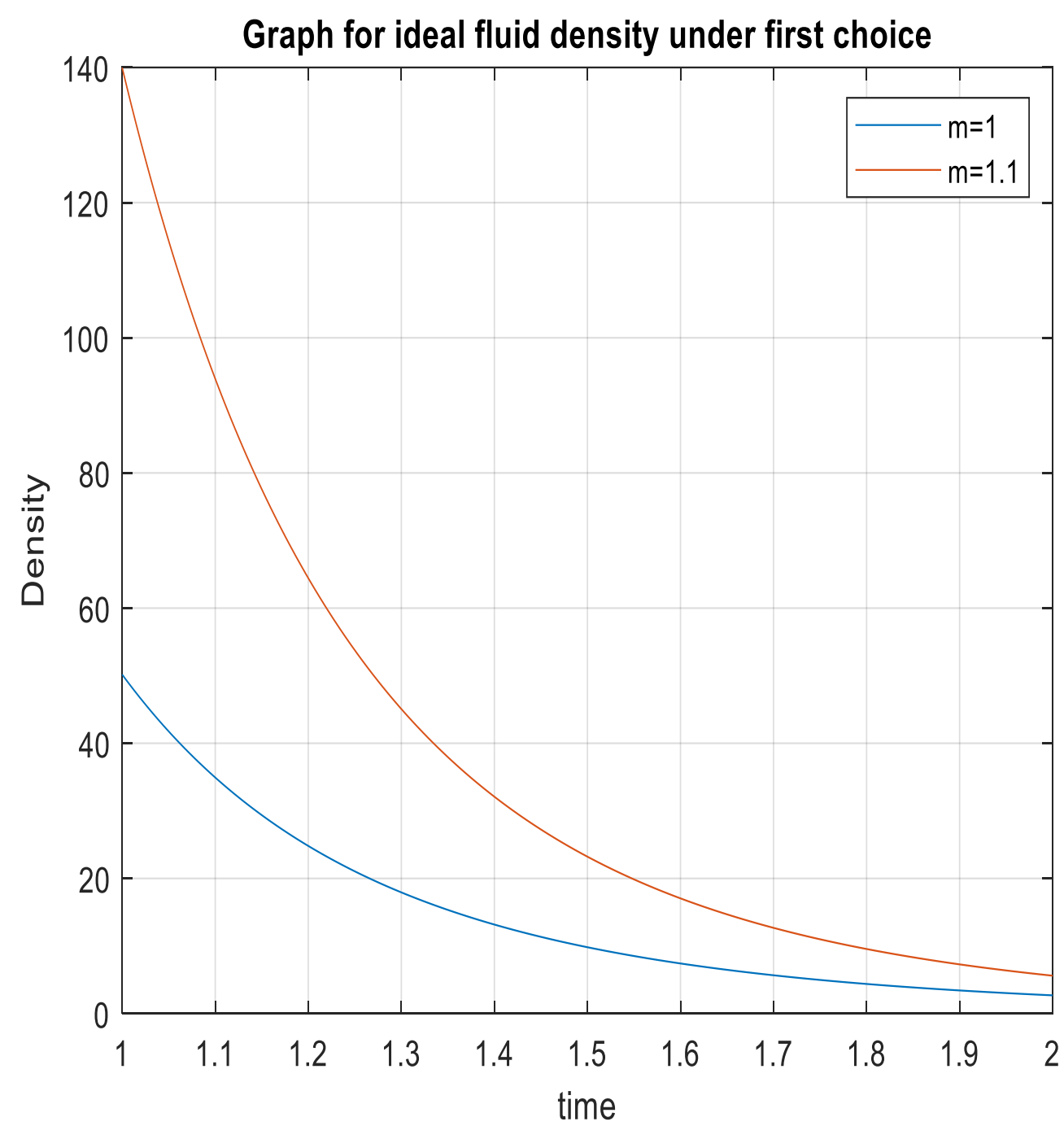

Figure 3: $\rho_d$ vs time

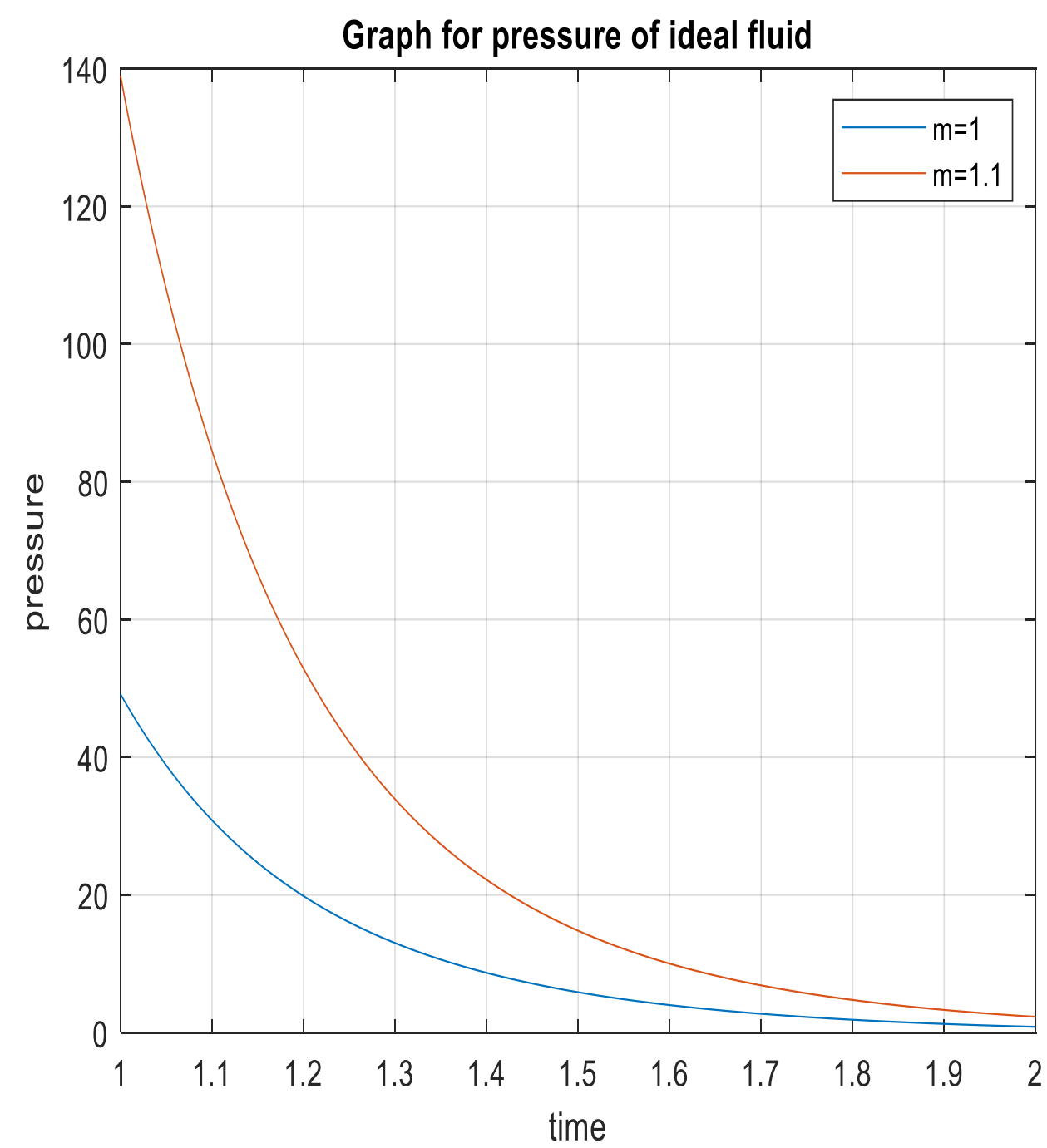

Figure 4: $p_d$ vs time

### 6. Derivations for ideal fluid energy density with Type II Inhomogeneous EOS

Now we'll derive $\rho_d$ using the inhomogeneous EOS for second case.

From our 2nd type of EOS we get;

$$\dot{\rho}_d + \frac{3mn\rho_d(1+\delta+A)}{(a_1+nt)} + \frac{3Bm^3n^3}{(a_1+nt)^3} = 0 \quad (33)$$

So, let's discuss about the solution of the choice of density of dark fluid from the above discussed differential equations. We get;

$$\rho_d = -\frac{3Bm^3n^3}{3mn(1+A+\delta)-2n}(a_1+nt)^{-2} + c(a_1+nt)^{-3m(1+A+\delta)} \quad (34)$$

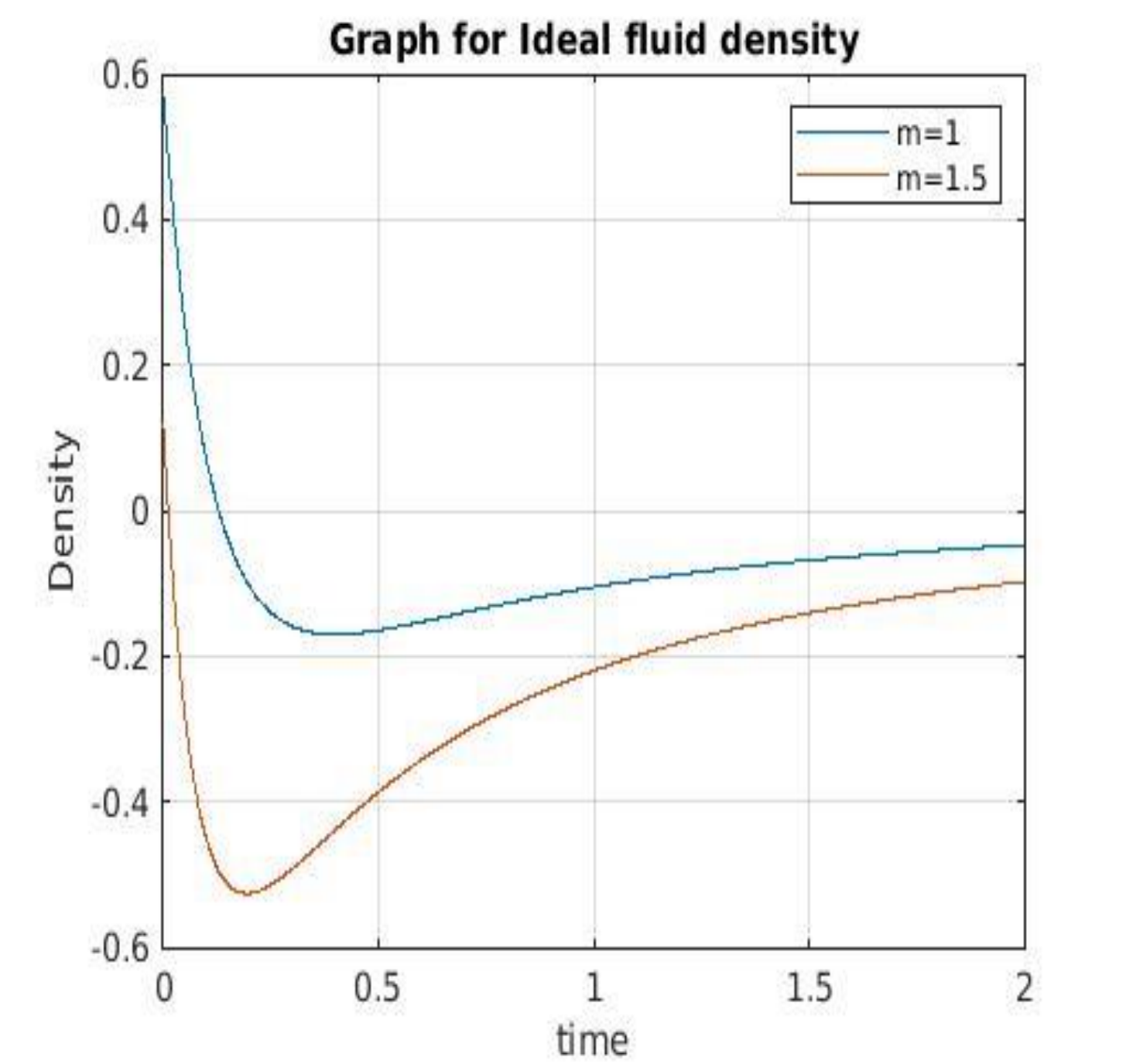

Figure 5: $\rho_d$ vs time

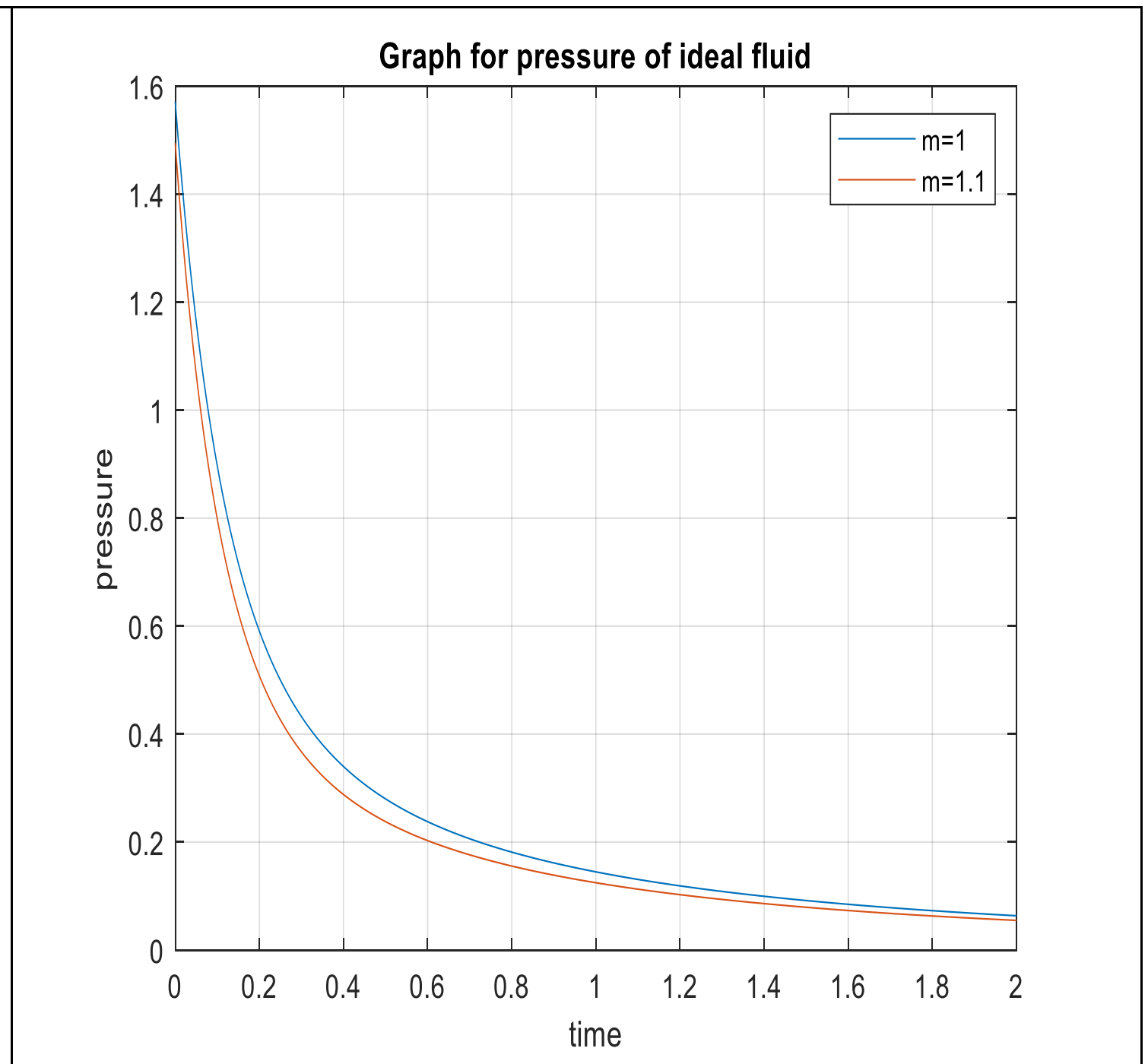

Figure 6: $p_d$ vs time

From the figure 5 we observe that the energy density for 2nd type EOS is becoming negative for higher values of m. Therefore, we can conclude that, this specific form of EOS is not stable for accelerating universe. It will be stable for decelerating universe for value m < 1.

From section 4 and 5 For type I, we can find positive pressure as well as positive density that only controls the fluid properties and thermodynamics of the system. Type I EOS also provides the stable fluid definition for both accelerated and decelerated expanding universe.

## 7. Calculations for Scalar field, Scalar field densities, Scalar field potentials and also gravitational constants for those above chosen Inhomogeneous EOS

Now, we get;

$$\rho_\varphi = \frac{3m^2n^2}{G(t)(a_1+nt)^2} - \rho_m - \rho_d - \rho_c - \frac{\Lambda(t)}{3} \; ; \qquad \text{where we have taken } 8\pi = 1 = c \qquad (35)$$

So, we get;

$$\varphi = \varphi_0 + \int \left[\frac{2}{G(t)}\left(\left(\frac{\ddot{a}}{a}\right) - \left(\frac{\dot{a}}{a}\right)^2\right) - \left((\rho_d + p_d) + (\rho_c + p_c) + (\rho_m + p_m)\right)\right]^{\frac{1}{2}} dt \qquad (36)$$

Again, we get;

$$V(\varphi) = \frac{2}{G(t)}\left(2\left(\frac{\dot{a}}{a}\right)^2 + \left(\frac{\ddot{a}}{a}\right)\right) - \left((\rho_d - p_d) + (\rho_c - p_c) + (\rho_m - p_m)\right) - 12\frac{\Lambda(t)}{G(t)} \qquad (37)$$

And

$$G(t) = \int -\frac{\dot{\Lambda}}{\rho_d} dt + G_0 \tag{38}$$

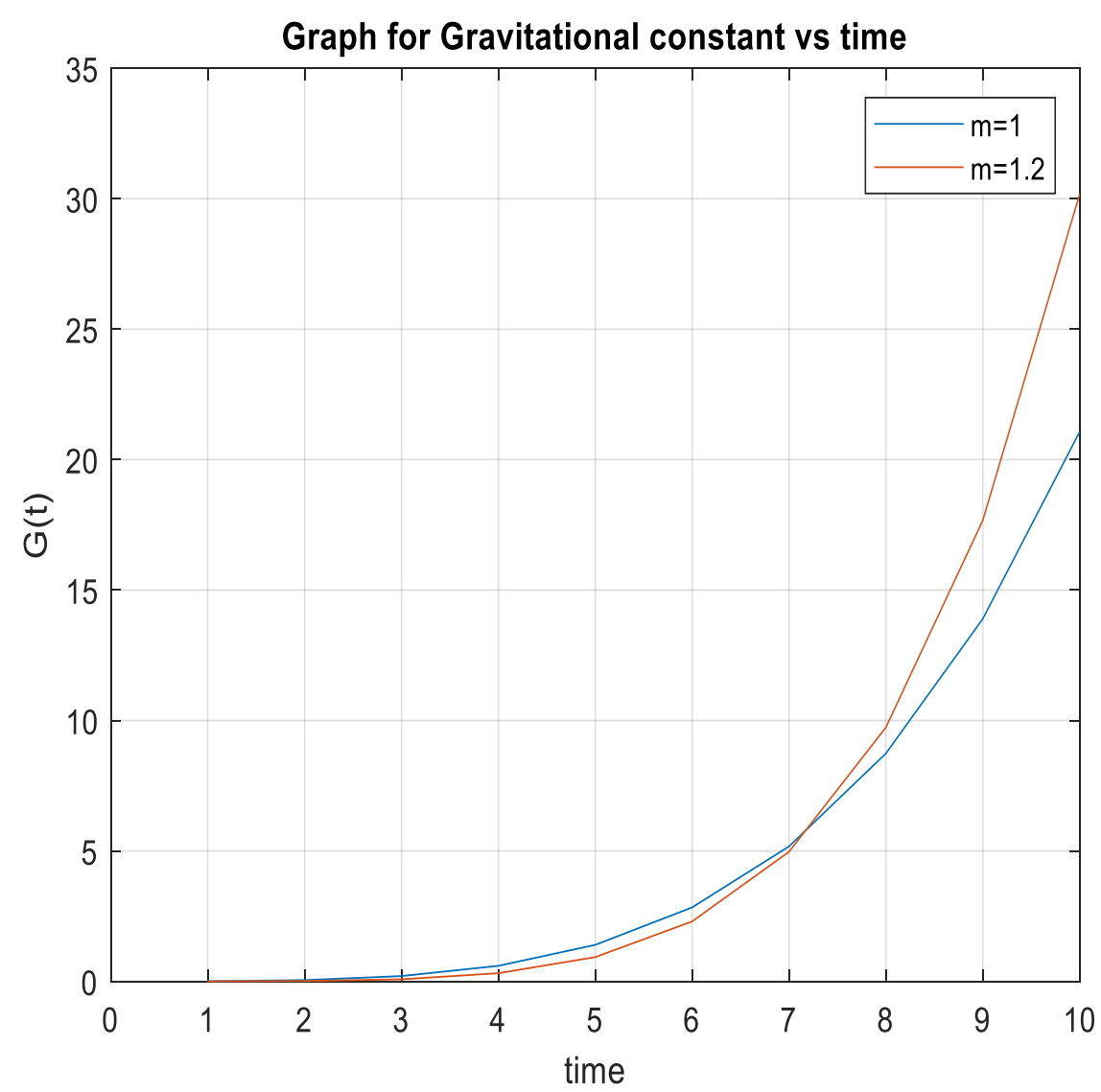

Figure 7: G(t) vs time for Type I

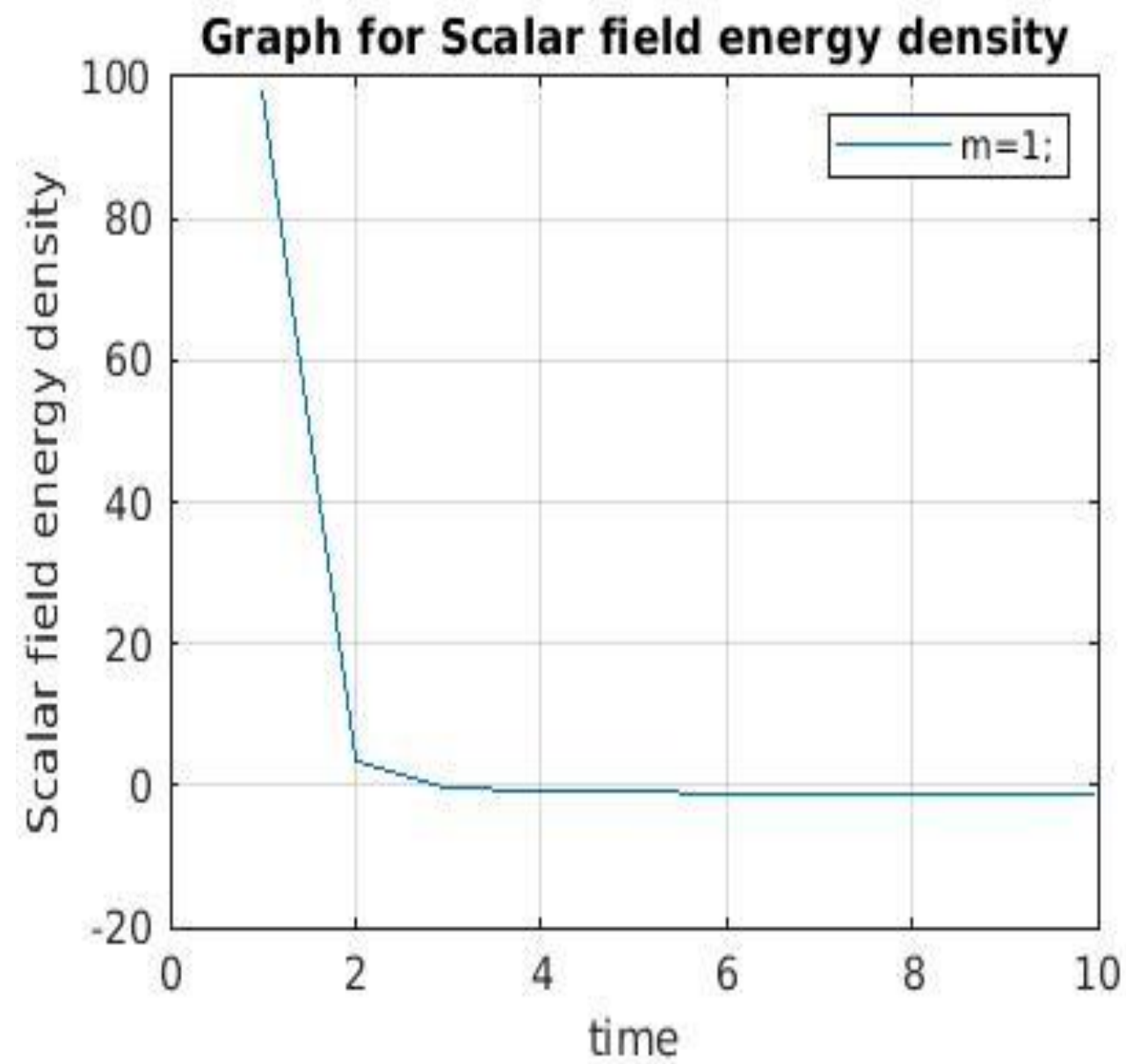

Figure 8: $\rho_\varphi$ vs time for Type I

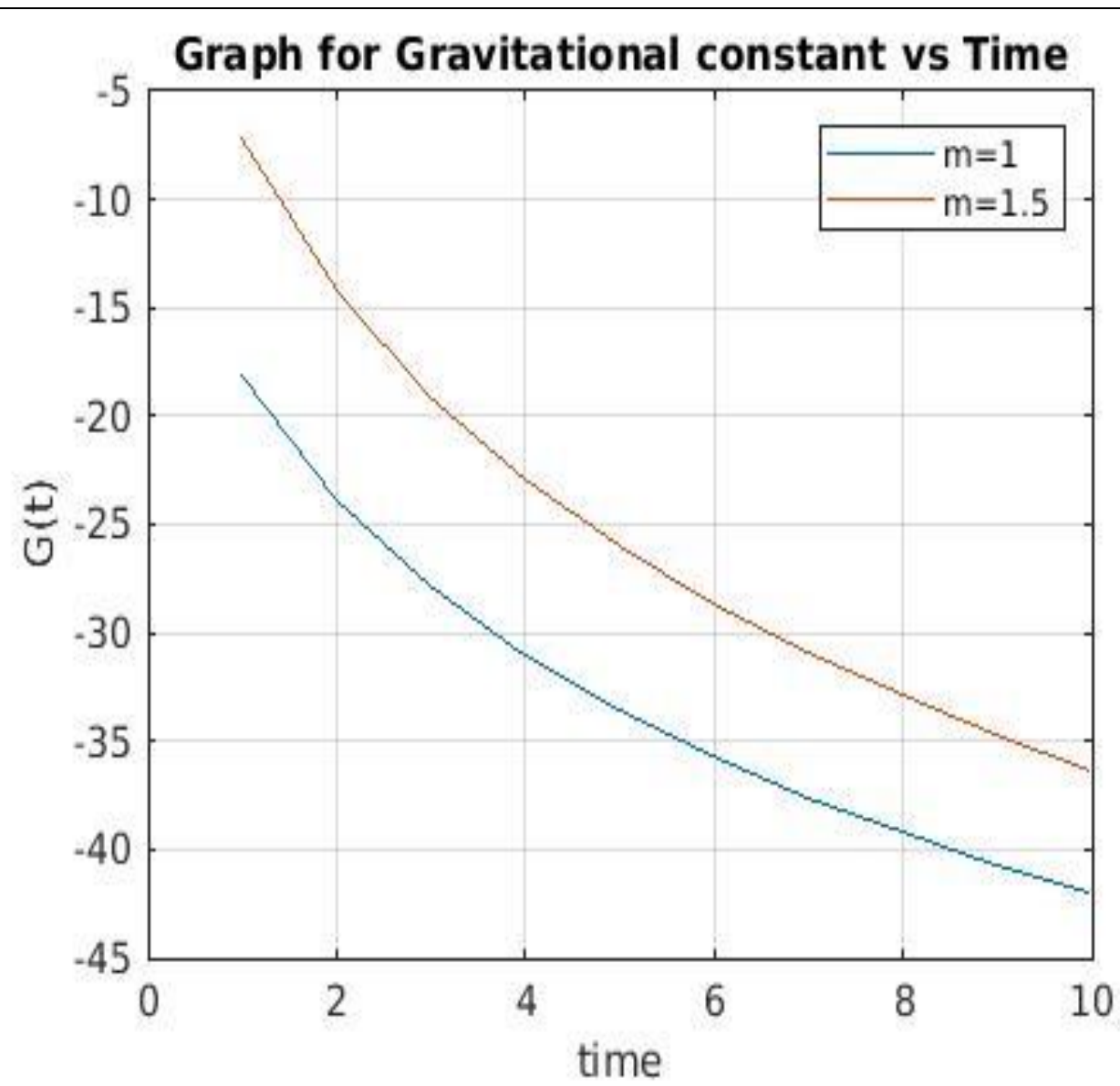

Figure 9: G(t) vs time for Type II

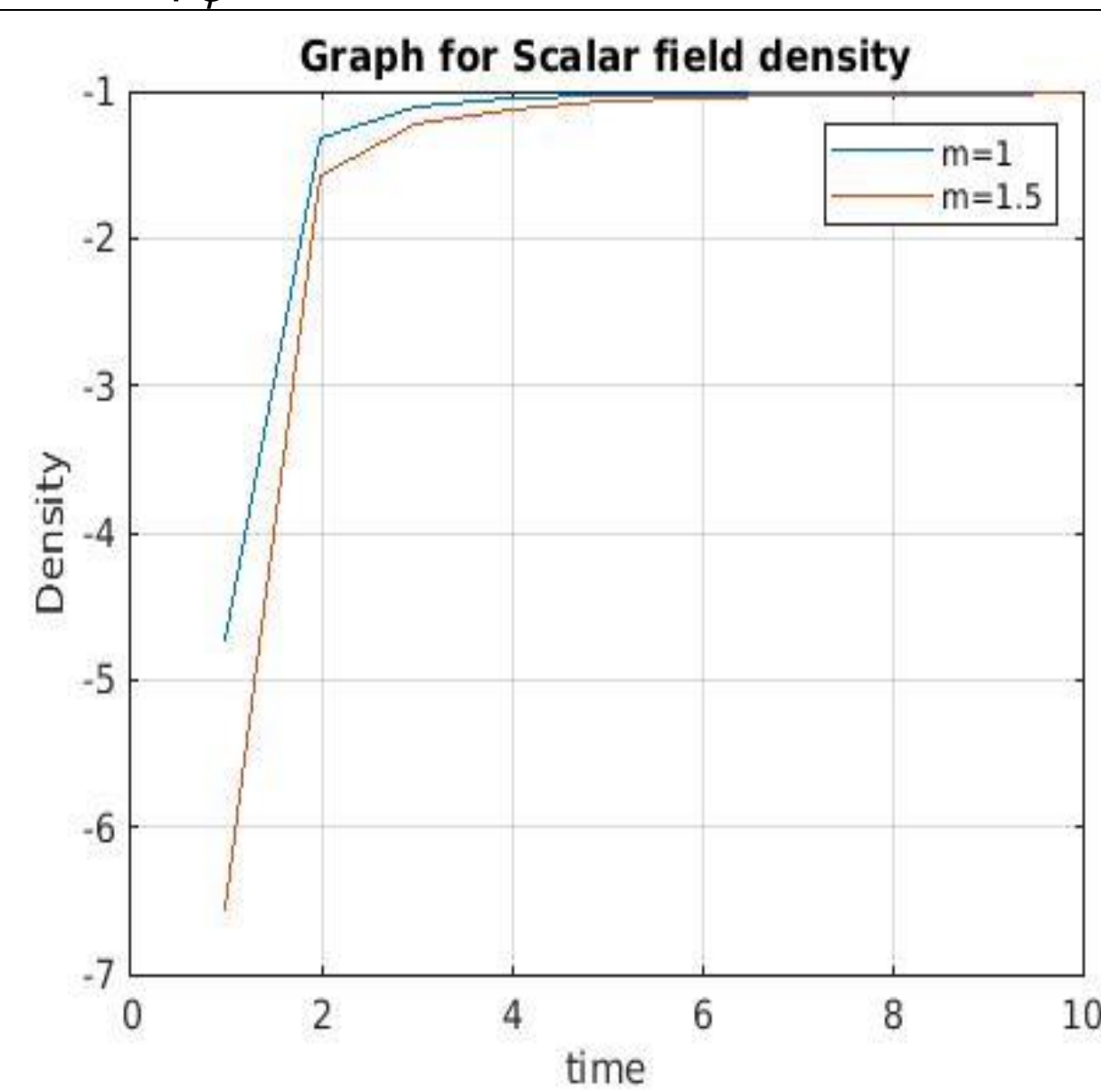

Figure 10: $\rho_\varphi$ vs time for Type II

For type I we can observe the value of G is almost zero in the initial time of the universe.

In figure 9 and 10 we observe that the value of G and scalar field density come out to be negative which leads to negative kinetic energy. Therefore, we can conclude that 2nd type of EOS is not stable for accelerating universe.

## 8. Analysis of Thermodynamics energy conditions in the purview of the above used case I Inhomogeneous EOS

From equation (3) we observe that the universe comprises of 4 components each of which has their role in the evolution of the universe. In this paper we investigate the effect of inhomogeneous fluid in the cosmic evolution. We assume that the fluid property of the universe is determined by the EOS of Inhomogeneous fluid. Therefore, we investigate the thermodynamics w.r.t energy density and pressure of that inhomogeneous fluid. (The phantom universe condition through EOS parameters are brought with both Inhomogeneous and Dark fluid.)

For WEC and NEC they have a common statement and that is $\rho_d + p_d \geq 0$. So, we have only used NEC.

From our derivations we see that the time varying function for scale factor, energy density and effective pressure can be expressed as follows.

$$a(t) = a_{01}(a_1 + nt)^m$$

$$\rho_d = c_1 \exp\left(-\frac{constant1*\log(a_1+nt) - \frac{constant2*(a_1+nt)^{-\alpha}}{\alpha}}{n}\right) + \frac{constant2*c}{a_2*\alpha*n} * \left(\exp\left(-\frac{constant1*\log(a_1+nt) - \frac{constant2*(a_1+nt)^{-\alpha}}{\alpha}}{n}\right)\right) * (a_1+nt)^{-\beta+\frac{constant1}{n}} * \left(\frac{constant2*(a_1+nt)^{-\alpha}}{\alpha*n}\right)^{\frac{-\beta+\frac{constant1}{n}}{\alpha}} Gamma\left(-\frac{-\beta+\frac{constant1}{n}}{\alpha}, \frac{constant2*(a_1+nt)^{-\alpha}}{\alpha*n}\right); \quad (39)$$

Here constant1 = $3mn(1+\delta)$; constant2 = $3mna_2$; $c_1$ = constant of integration.

And

$$p_d = a_2 t^{-\alpha}\left[c_1 \exp\left(-\frac{constant1*\log(a_1+nt) - \frac{constant2*(a_1+nt)^{-\alpha}}{\alpha}}{n}\right) + \frac{constant2*c}{a_2*\alpha*n} * \left(\exp\left(-\frac{constant1*\log(a_1+nt) - \frac{constant2*(a_1+nt)^{-\alpha}}{\alpha}}{n}\right)\right) * (a_1+nt)^{-\beta+\frac{constant1}{n}} * \left(\frac{constant2*(a_1+nt)^{-\alpha}}{\alpha*n}\right)^{\frac{-\beta+\frac{constant1}{n}}{\alpha}} Gamma\left(-\frac{-\beta+\frac{constant1}{n}}{\alpha}, \frac{constant2*(a_1+nt)^{-\alpha}}{\alpha*n}\right)\right] - ct^{-\beta} \quad (40)$$

So we observe that none of the above will show finite time singularity. Therefore the finite time future singularity problems has now been resolved using the type I case I Inhomogeneous EOS.

Now we get from above expressions;

$$\rho_d + p_d = \{c_1 \exp\left(-\frac{constant1*\log(a_1+nt) - \frac{constant2*(a_1+nt)^{-\alpha}}{\alpha}}{n}\right) + \frac{constant2*c}{a_2*\alpha*n} *$$

$$(\exp\left(-\frac{constant1*\log(a_1+nt) - \frac{constant2*(a_1+nt)^{-\alpha}}{\alpha}}{n}\right)) * (a_1+nt)^{-\beta+\frac{constant1}{n}} *$$

$$\left(\frac{constant2*(a_1+nt)^{-\alpha}}{\alpha*n}\right)^{\frac{-\beta+\frac{constant1}{n}}{\alpha}} Gamma\left(-\frac{-\beta+\frac{constant1}{n}}{\alpha}, \frac{constant2*(a_1+nt)^{-\alpha}}{\alpha*n}\right)\} +$$

$$\{a_2 t^{-\alpha}[c_1 \exp\left(-\frac{constant1*\log(a_1+nt) - \frac{constant2*(a_1+nt)^{-\alpha}}{\alpha}}{n}\right) + \frac{constant2*c}{a_2*\alpha*n} *$$

$$(\exp\left(-\frac{constant1*\log(a_1+nt) - \frac{constant2*(a_1+nt)^{-\alpha}}{\alpha}}{n}\right)) * (a_1+nt)^{-\beta+\frac{constant1}{n}} *$$

$$\left(\frac{constant2*(a_1+nt)^{-\alpha}}{\alpha*n}\right)^{\frac{-\beta+\frac{constant1}{n}}{\alpha}} Gamma(-\frac{-\beta+\frac{constant1}{n}}{\alpha}, \frac{constant2*(a_1+nt)^{-\alpha}}{\alpha*n})\,] - ct^{-\beta}\} \tag{41}$$

And

$$\rho_d + 3p_d = \{c_1 \exp\left(-\frac{constant1*\log(a_1+nt) - \frac{constant2*(a_1+nt)^{-\alpha}}{\alpha}}{n}\right) + \frac{constant2*c}{a_2*\alpha*n} *$$

$$(\exp\left(-\frac{constant1*\log(a_1+nt) - \frac{constant2*(a_1+nt)^{-\alpha}}{\alpha}}{n}\right)) * (a_1+nt)^{-\beta+\frac{constant1}{n}} *$$

$$\left(\frac{constant2*(a_1+nt)^{-\alpha}}{\alpha*n}\right)^{\frac{-\beta+\frac{constant1}{n}}{\alpha}} Gamma\left(-\frac{-\beta+\frac{constant1}{n}}{\alpha}, \frac{constant2*(a_1+nt)^{-\alpha}}{\alpha*n}\right)\} +$$

$$3\{a_2 t^{-\alpha}[c_1 \exp\left(-\frac{constant1*\log(a_1+nt) - \frac{constant2*(a_1+nt)^{-\alpha}}{\alpha}}{n}\right) + \frac{constant2*c}{a_2*\alpha*n} *$$

$$(\exp\left(-\frac{constant1*\log(a_1+nt) - \frac{constant2*(a_1+nt)^{-\alpha}}{\alpha}}{n}\right)) * (a_1+nt)^{-\beta+\frac{constant1}{n}} *$$

$$\left(\frac{constant2*(a_1+nt)^{-\alpha}}{\alpha*n}\right)^{\frac{-\beta+\frac{constant1}{n}}{\alpha}} Gamma(-\frac{-\beta+\frac{constant1}{n}}{\alpha}, \frac{constant2*(a_1+nt)^{-\alpha}}{\alpha*n})\,] - ct^{-\beta}\} \tag{42}$$

For DEC we need to get following condition.

$$\rho_d - p_d = \{c_1 \exp\left(-\frac{constant1*\log(a_1+nt) - \frac{constant2*(a_1+nt)^{-\alpha}}{\alpha}}{n}\right) + \frac{constant2*c}{a_2*\alpha*n} *$$

$$(\exp\left(-\frac{constant1*\log(a_1+nt) - \frac{constant2*(a_1+nt)^{-\alpha}}{\alpha}}{n}\right)) * (a_1+nt)^{-\beta+\frac{constant1}{n}} *$$

$$\left(\frac{constant2*(a_1+nt)^{-\alpha}}{\alpha*n}\right)^{\frac{-\beta+\frac{constant1}{n}}{\alpha}} Gamma\left(-\frac{-\beta+\frac{constant1}{n}}{\alpha}, \frac{constant2*(a_1+nt)^{-\alpha}}{\alpha*n}\right)\} +$$

$$\{a_2 t^{-\alpha}[c_1 \exp\left(-\frac{constant1*\log(a_1+nt) - \frac{constant2*(a_1+nt)^{-\alpha}}{\alpha}}{n}\right) + \frac{constant2*c}{a_2*\alpha*n} *$$

$$(\exp\left(-\frac{constant1*\log(a_1+nt) - \frac{constant2*(a_1+nt)^{-\alpha}}{\alpha}}{n}\right)) * (a_1+nt)^{-\beta+\frac{constant1}{n}} *$$

$$\left(\frac{constant2*(a_1+nt)^{-\alpha}}{\alpha*n}\right)^{\frac{-\beta+\frac{constant1}{n}}{\alpha}} Gamma(-\frac{-\beta+\frac{constant1}{n}}{\alpha}, \frac{constant2*(a_1+nt)^{-\alpha}}{\alpha*n})\,] - ct^{-\beta}\} \geq 0 \tag{43}$$

Now to satisfy the energy conditions to fulfill the attractive gravity nature we may get the following curves.

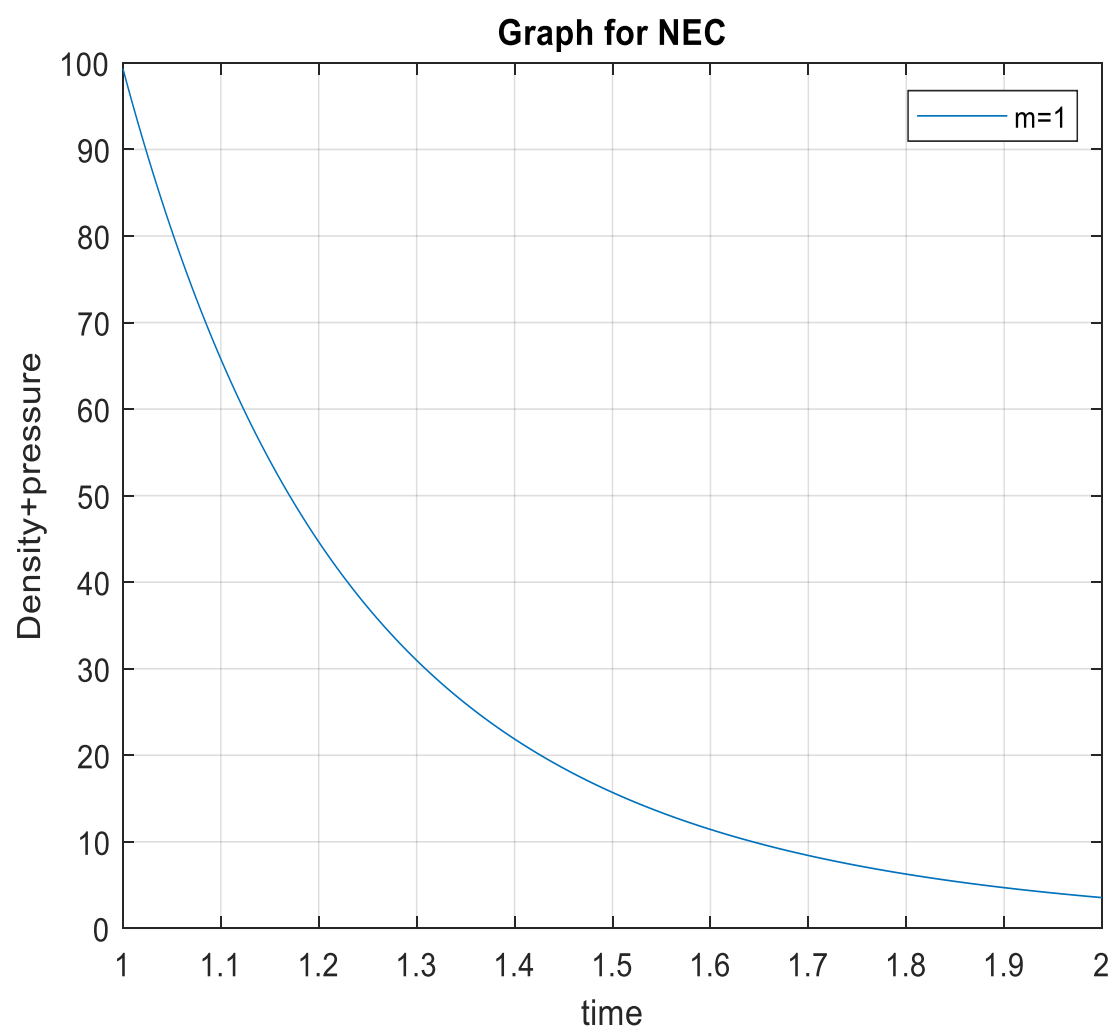

Figure 11: NEC vs time for Type I

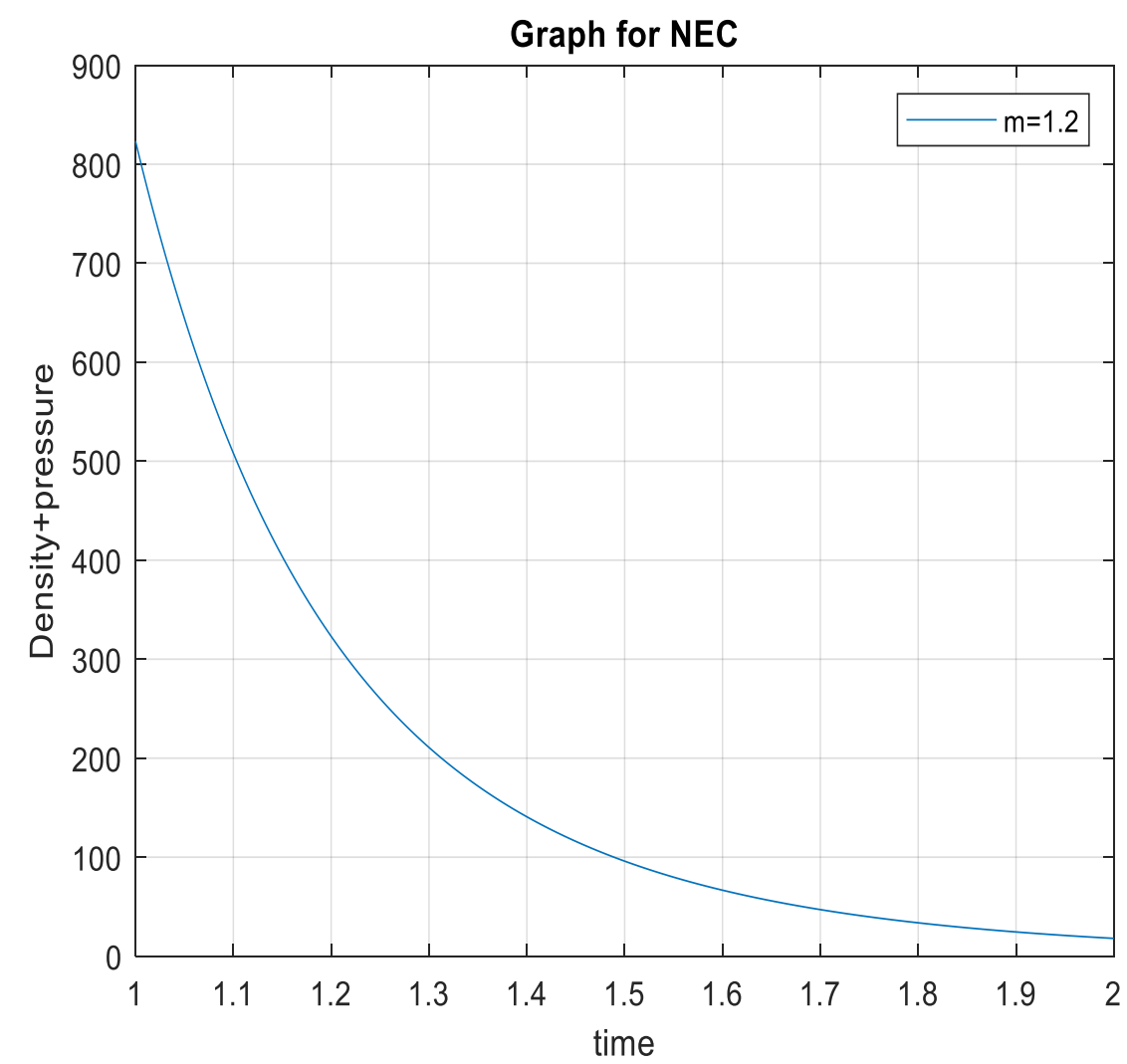

Figure 12: NEC vs time for type I

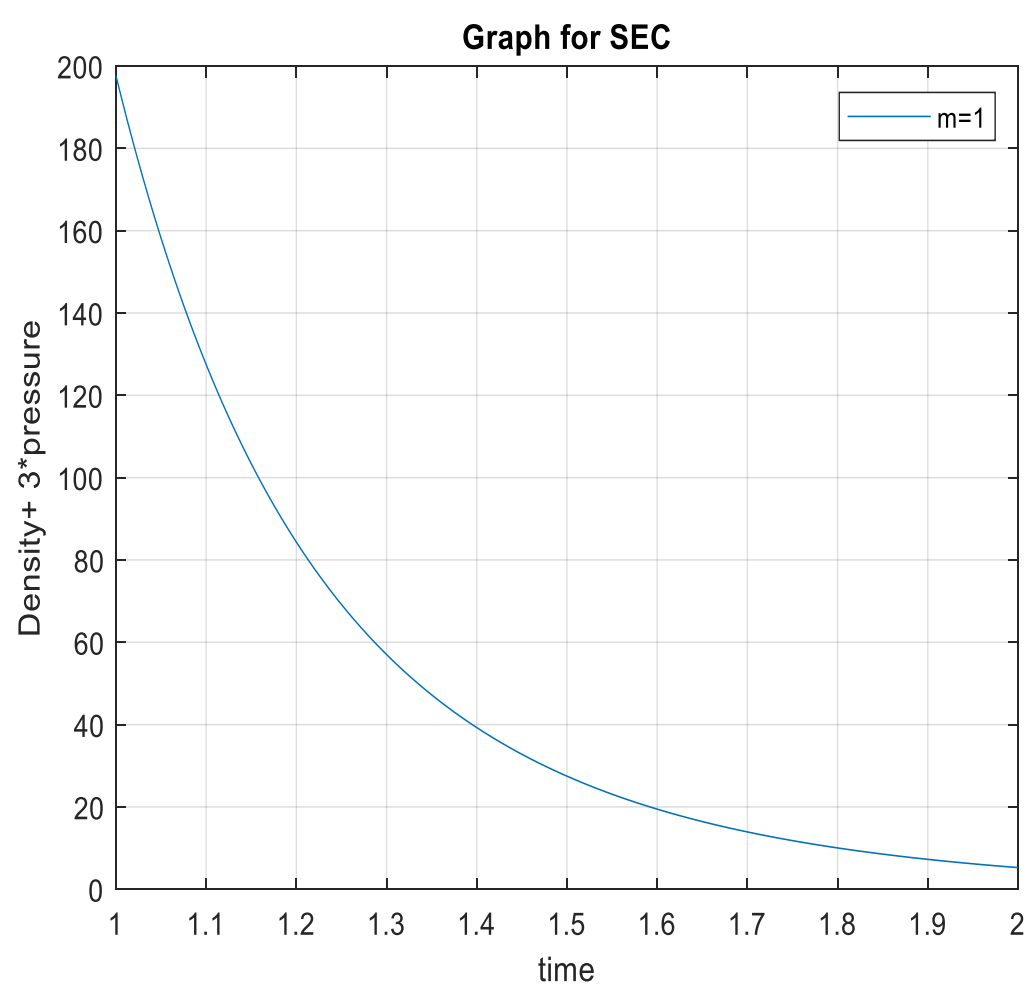

Figure 13: SEC vs time for Type I

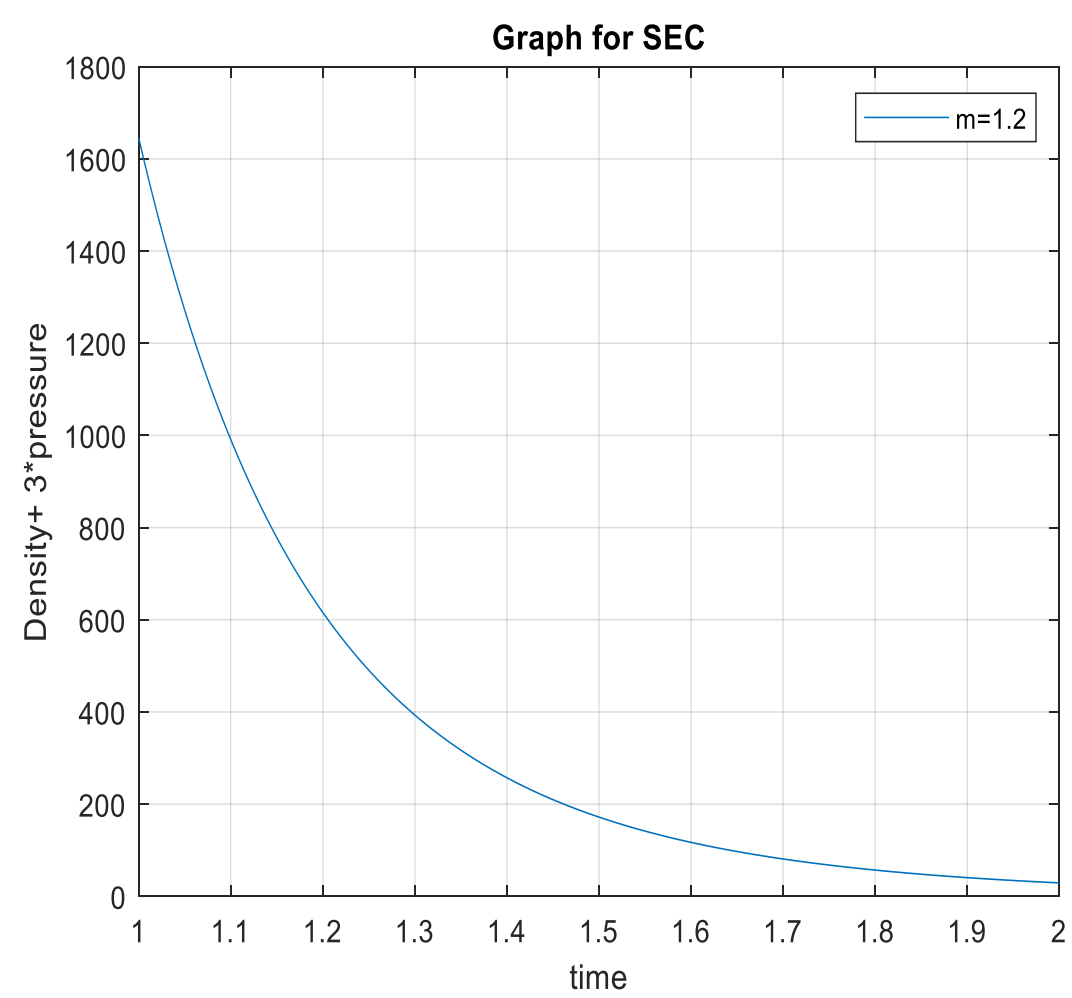

Figure 14: SEC vs time for type I

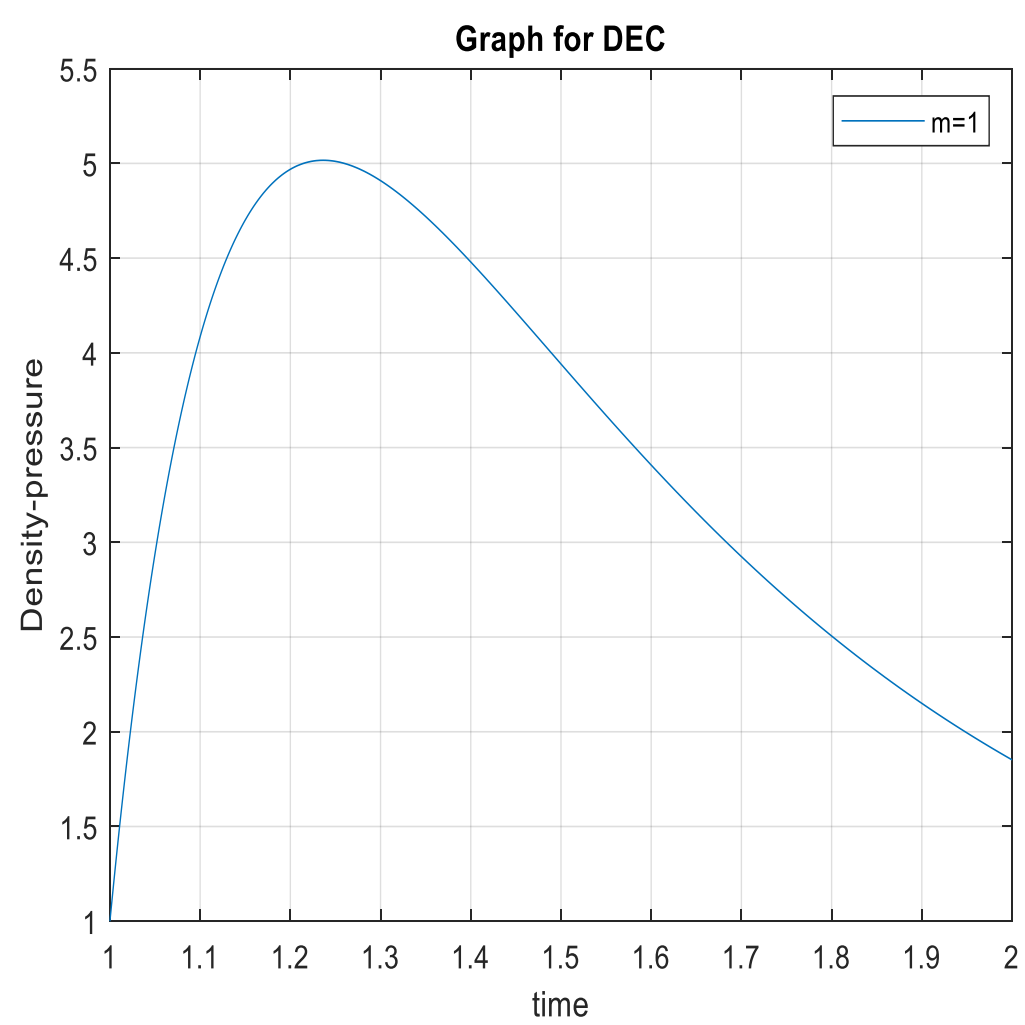

Figure 15: DEC vs time for Type I

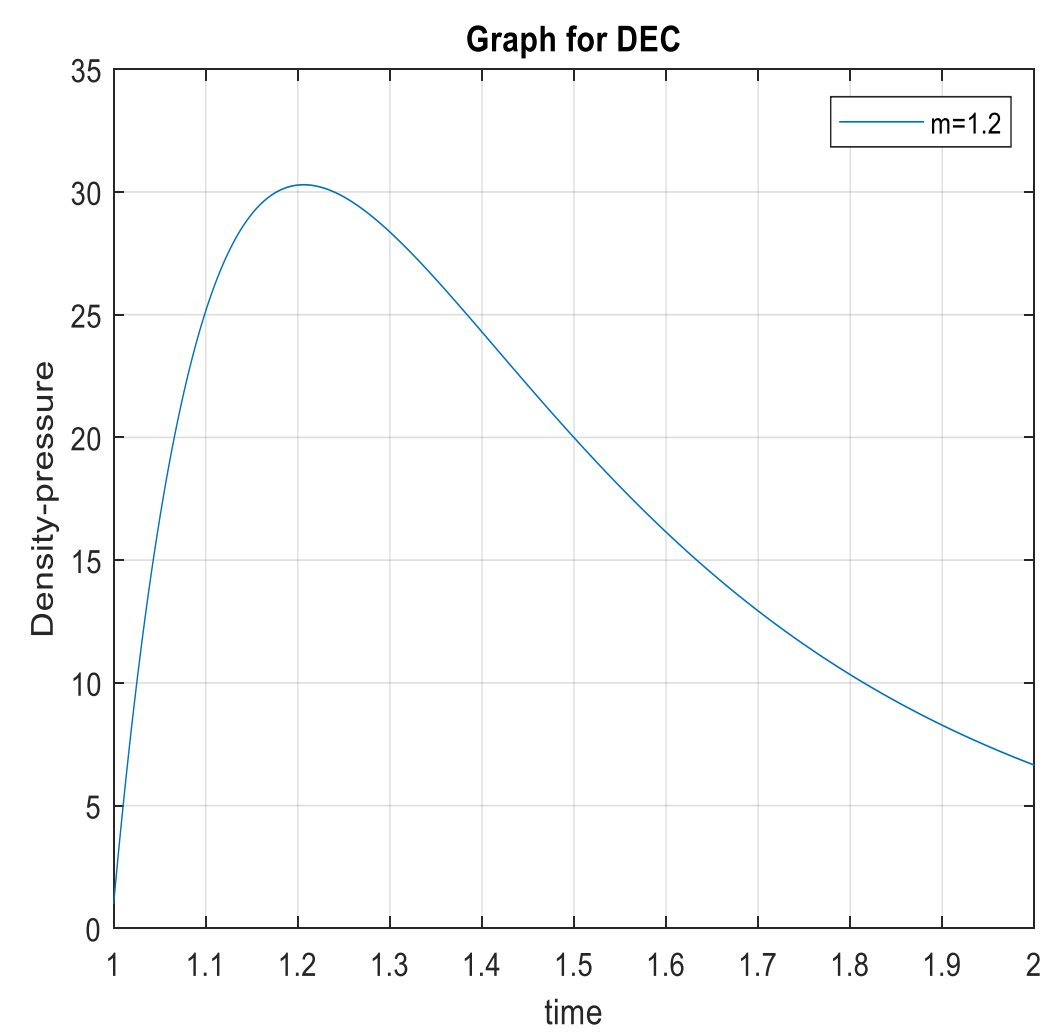

Figure 16: DEC vs time for type I

Therefore, we observe all the energy conditions satisfies with our calculated model of Inhomogeneous fluid.

## 9. Analysis of Thermodynamics energy conditions in the purview of the above used case II Inhomogeneous EOS

The basic assumptions behind multi-fluid system has been considered to be similar as discussed in section 7. The singularity problems have also been solved in similar ways.

From our derivations we see that the time varying function for scale factor, energy density and effective pressure can be expressed as follows.

$$a(t) = a_{01}(a_1 + nt)^m$$

$$\rho_d = -\frac{3Bm^3n^3}{3mn(1+A+\delta)-2n}(a_1+nt)^{-2} + c(a_1+nt)^{-3m(1+A+\delta)} \tag{44}$$

And

$$p_d = -A\frac{3Bm^3n^3}{3mn(1+A+\delta)-2n}(a_1+nt)^{-2} + c(a_1+nt)^{-3m(1+A+\delta)} + B\left(\frac{constant}{a_1+nt}\right)^2 \tag{45}$$

So we observe that none of the above will show finite time singularity. Therefore the finite time future singularity problems has now been resolved using the type II Inhomogeneous EOS.

Now we get from above expressions;

$$\rho_d+p_d = \left(-\frac{3Bm^3n^3}{3mn(1+A+\delta)-2n}(a_1+nt)^{-2} + c(a_1+nt)^{-3m(1+A+\delta)}\right) + \left(-A\frac{3Bm^3n^3}{3mn(1+A+\delta)-2n}(a_1+nt)^{-2} + c(a_1+nt)^{-3m(1+A+\delta)} + B\left(\frac{constant}{a_1+nt}\right)^2\right) \tag{46}$$

And

$$\rho_d+3p_d = \left(-\frac{3Bm^3n^3}{3mn(1+A+\delta)-2n}(a_1+nt)^{-2} + c(a_1+nt)^{-3m(1+A+\delta)}\right) + 3\left(-A\frac{3Bm^3n^3}{3mn(1+A+\delta)-2n}(a_1+nt)^{-2} + c(a_1+nt)^{-3m(1+A+\delta)} + B\left(\frac{constant}{a_1+nt}\right)^2\right) \tag{47}$$

For DEC we need to get following condition.

$$\rho_d-p_d = \left(-\frac{3Bm^3n^3}{3mn(1+A+\delta)-2n}(a_1+nt)^{-2} + c(a_1+nt)^{-3m(1+A+\delta)}\right) - \left(-A\frac{3Bm^3n^3}{3mn(1+A+\delta)-2n}(a_1+nt)^{-2} + c(a_1+nt)^{-3m(1+A+\delta)} + B\left(\frac{constant}{a_1+nt}\right)^2\right) \geq 0 \tag{48}$$

Now to satisfy the energy conditions to fulfill the attractive gravity nature we may get the following curves.

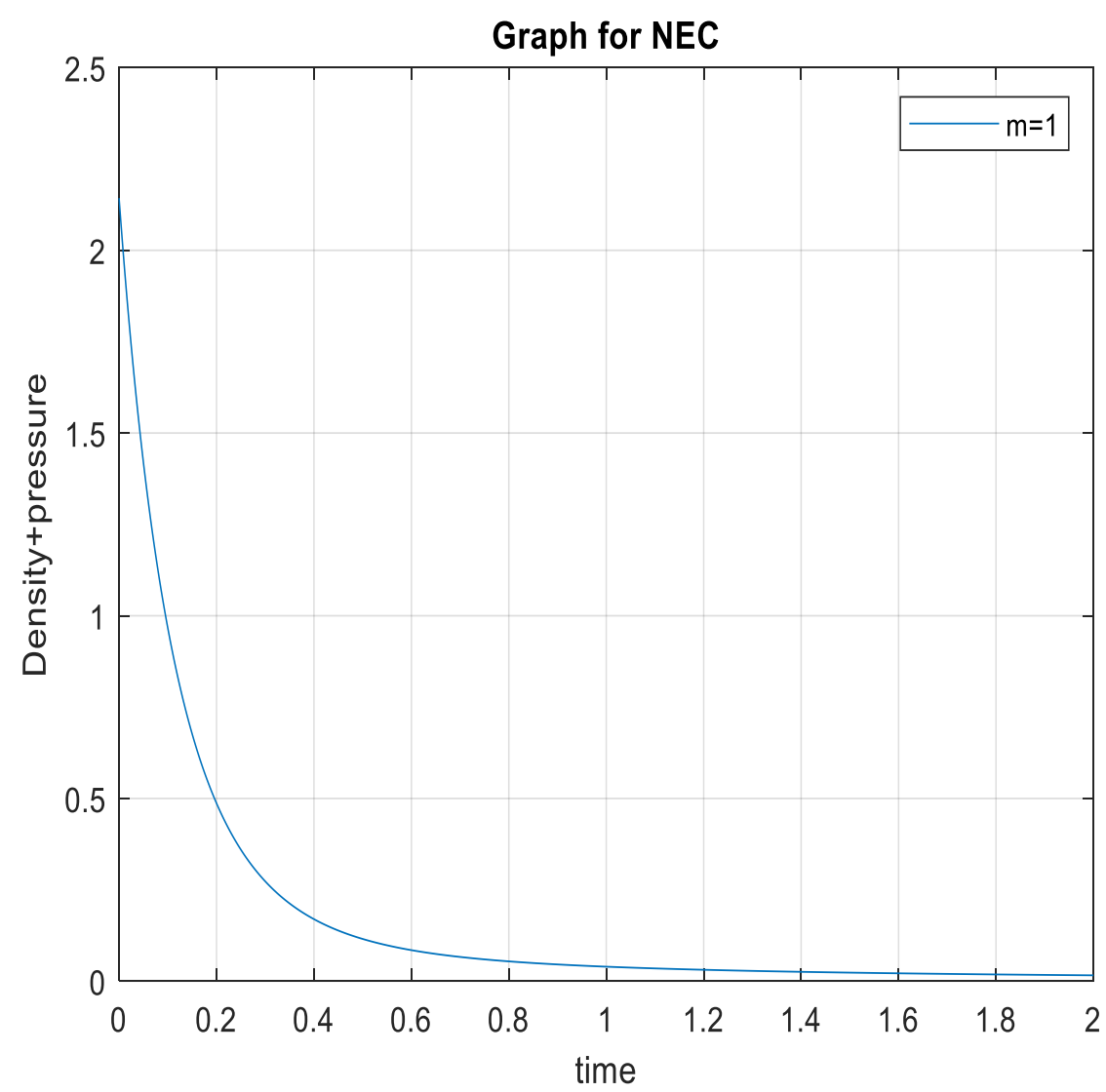

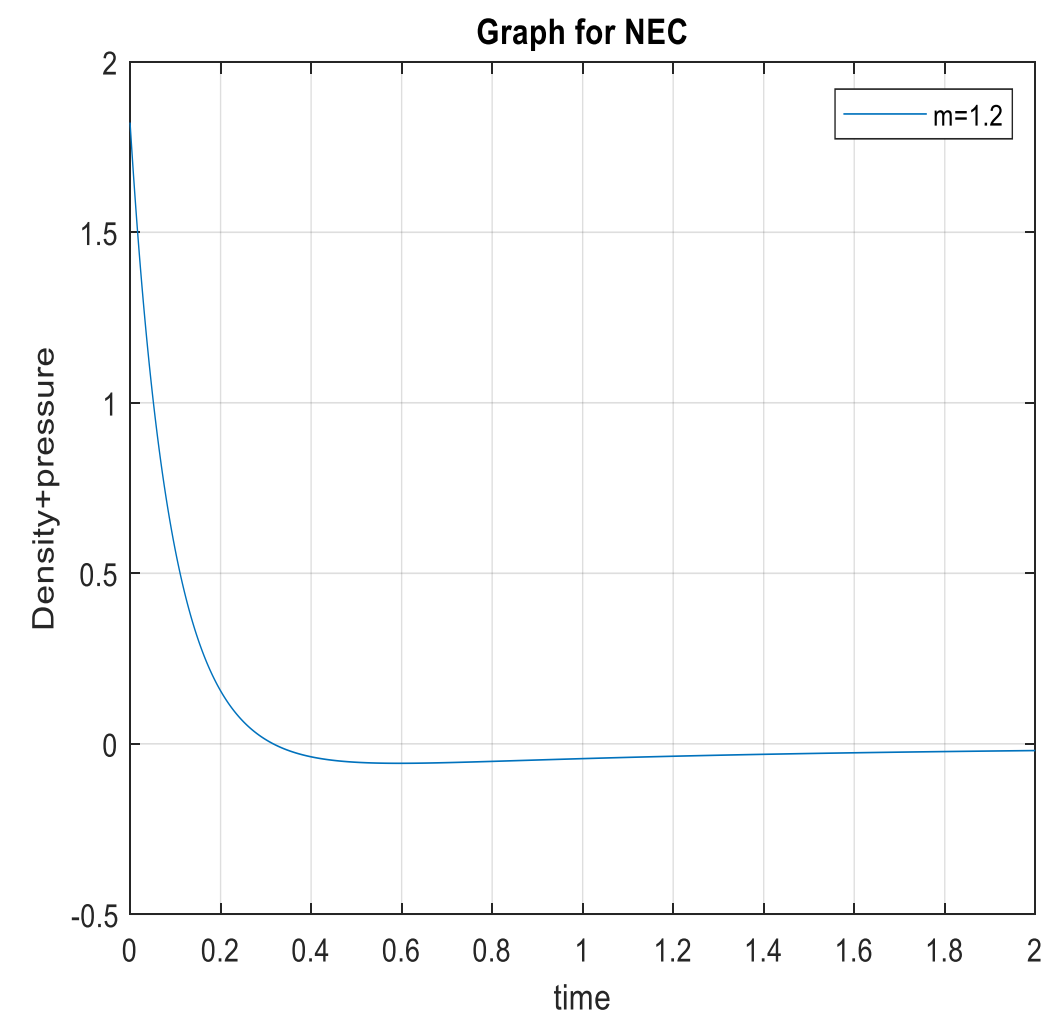

Figure 17: NEC vs time for Type II

Figure 18: NEC vs time for type II

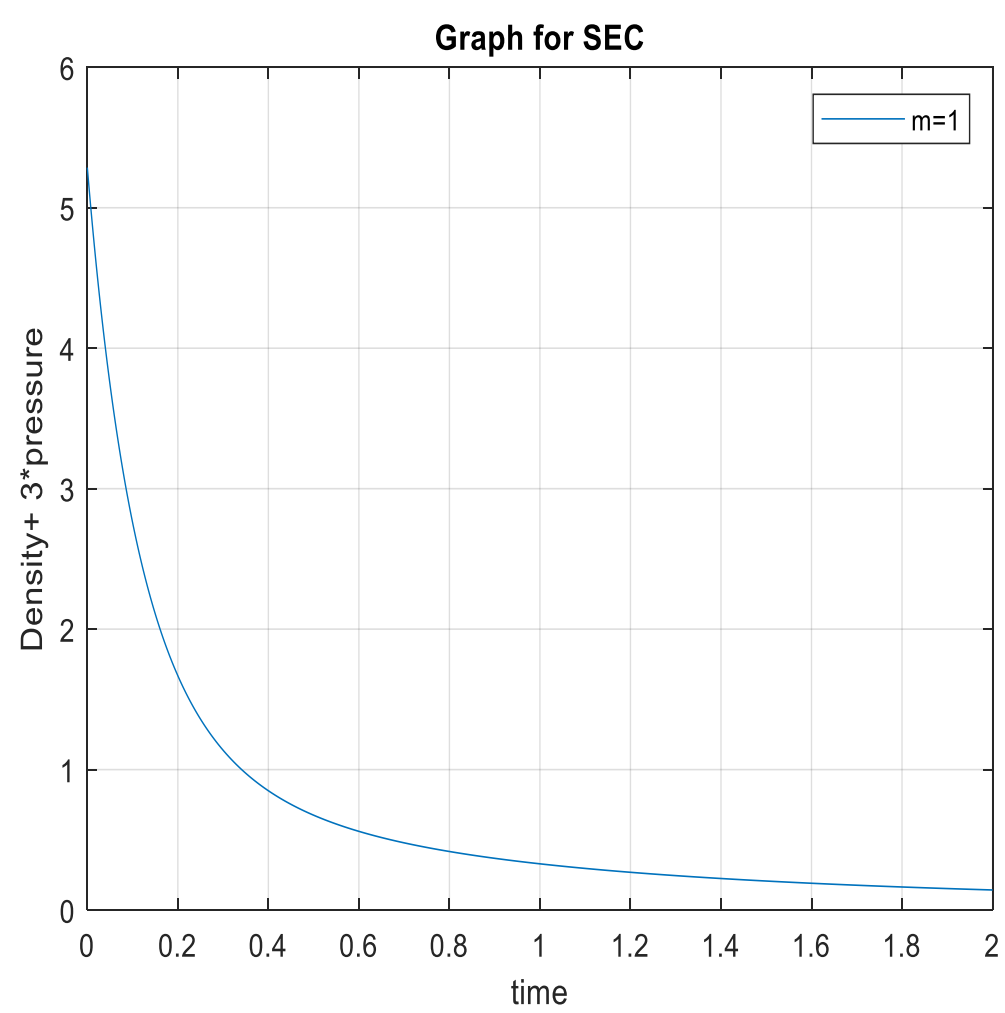

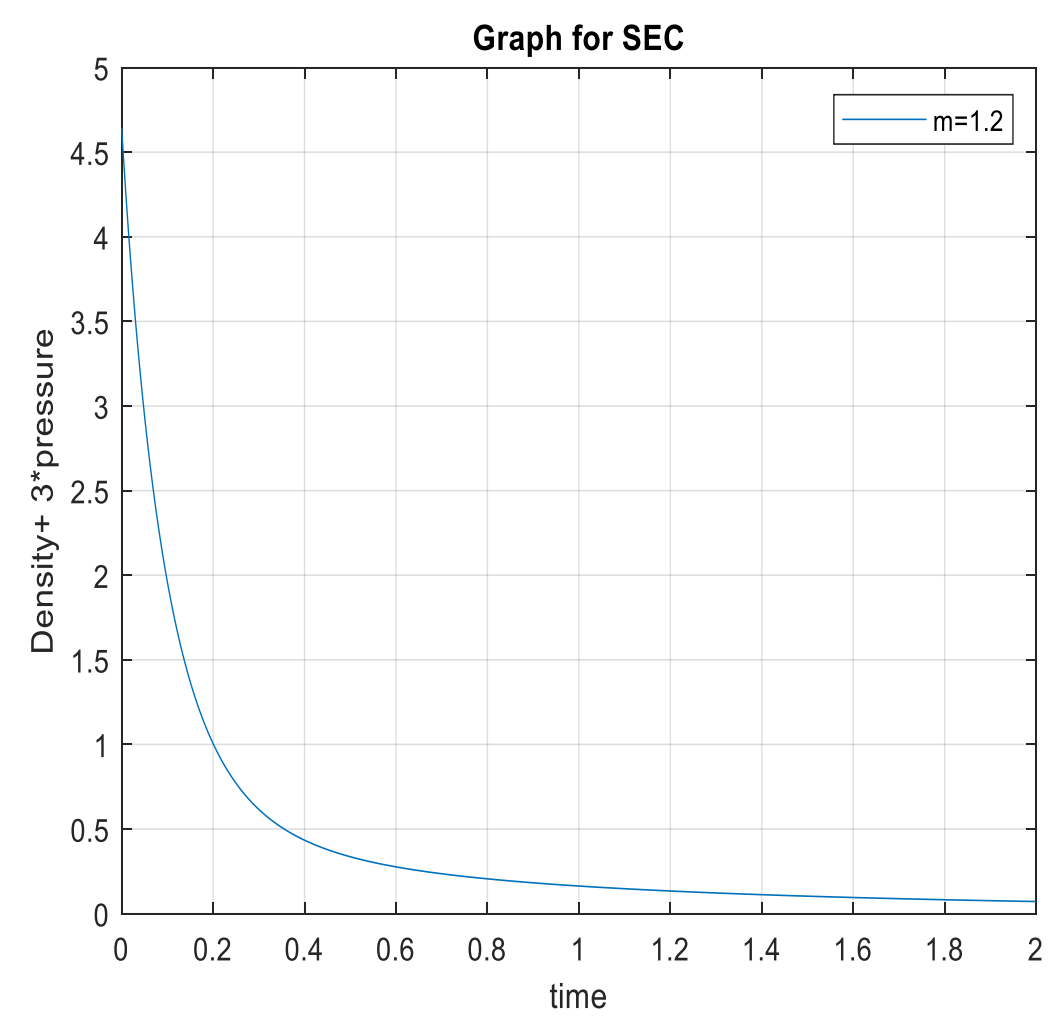

Figure 19: SEC vs time for Type II

Figure 20: SEC vs time for type II

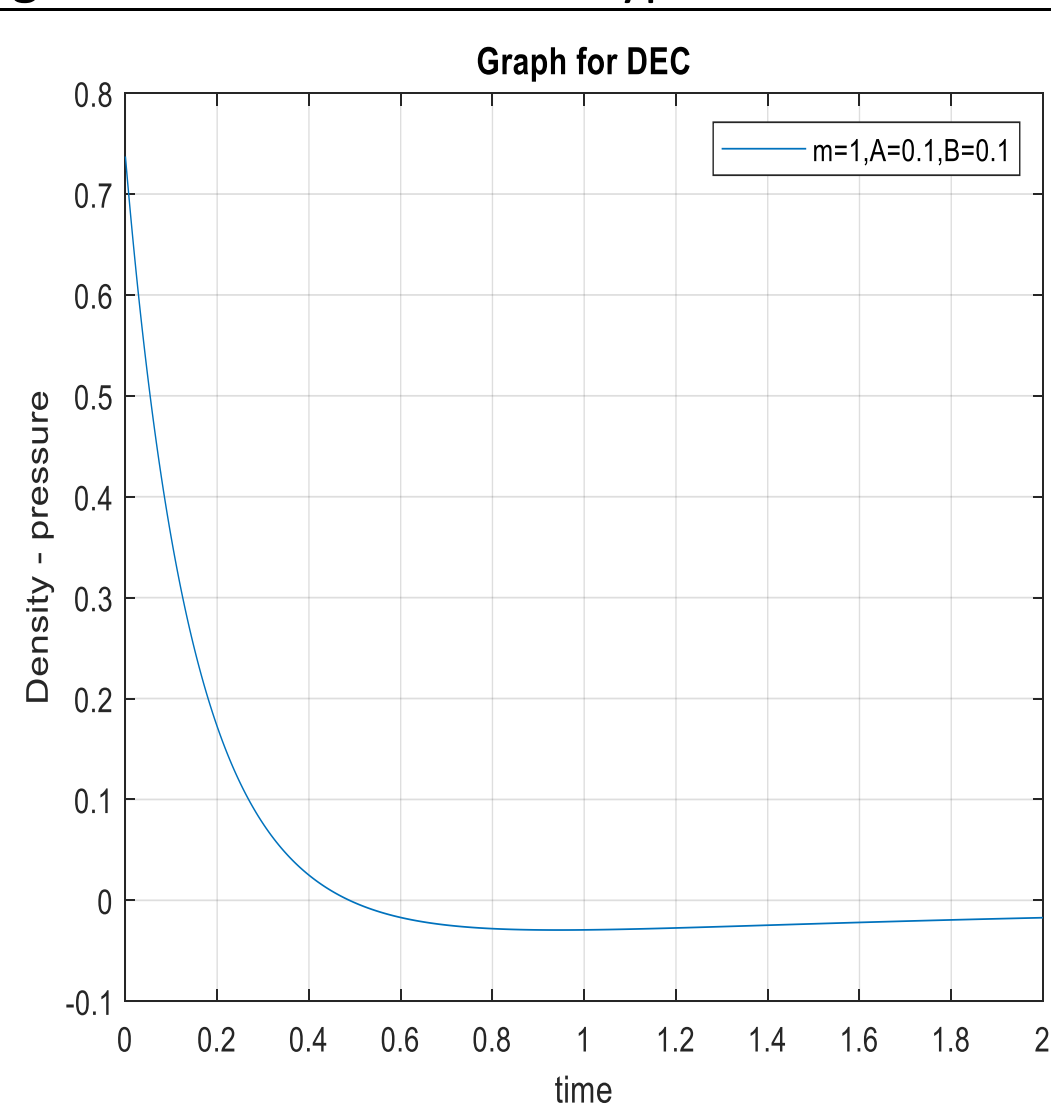

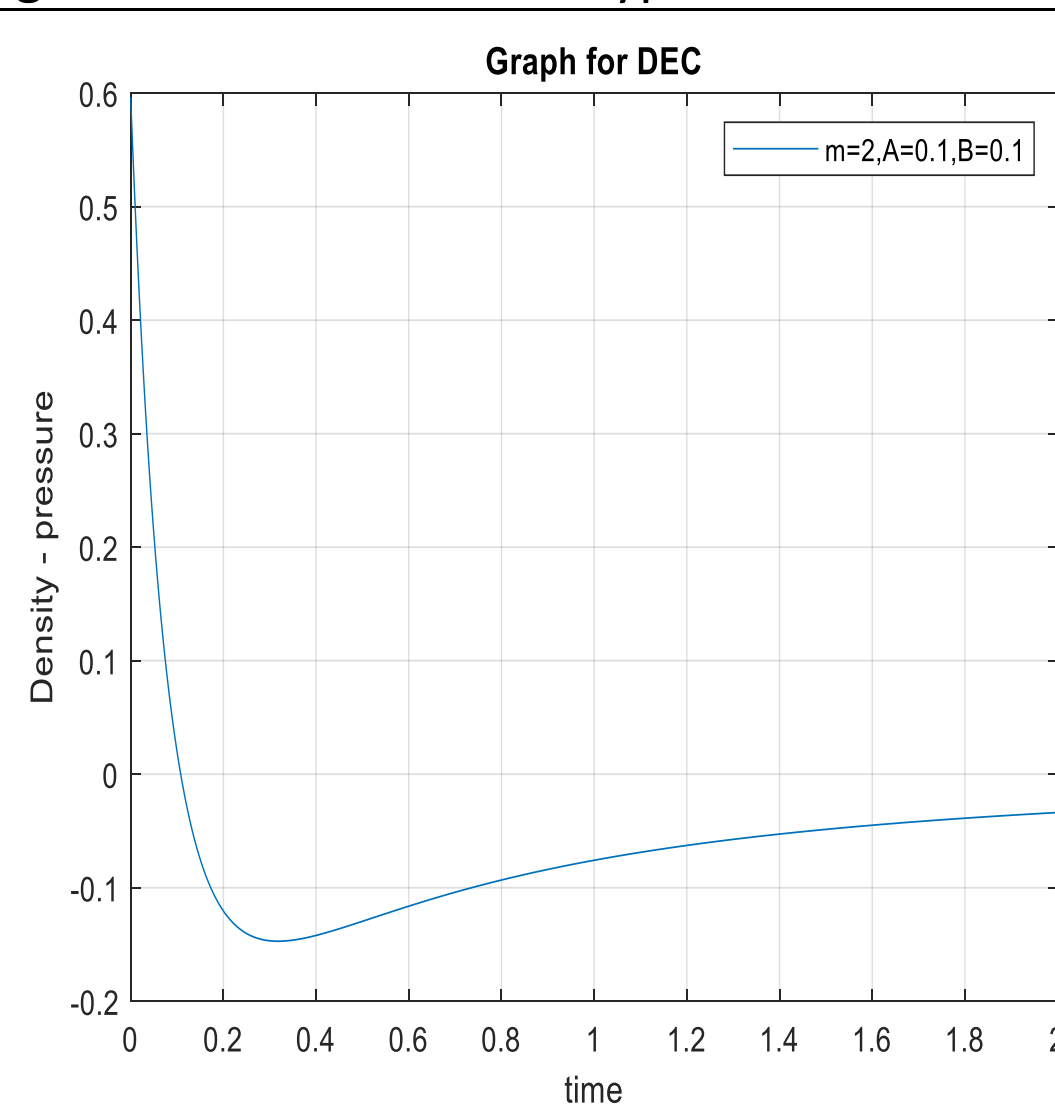

| Figure 21: DEC vs time for Type II | Figure 22: DEC vs time for type II |
|---|---|

For certain values of $m \geq 1$ (Accelerating universe) we see that WEC, NEC as well as DEC are not satisfied for $2^{nd}$ type of EOS. For m < 1 i.e. for decelerating universe all the energy conditions are satisfied.

Here solely the Inhomogeneous EOS parameter can provide the condition for phantom universe and that is why we solve the future time singularity problems with the energy density of inhomogeneous fluid.

## 10. Analysis of finite time future singularity problems in this model

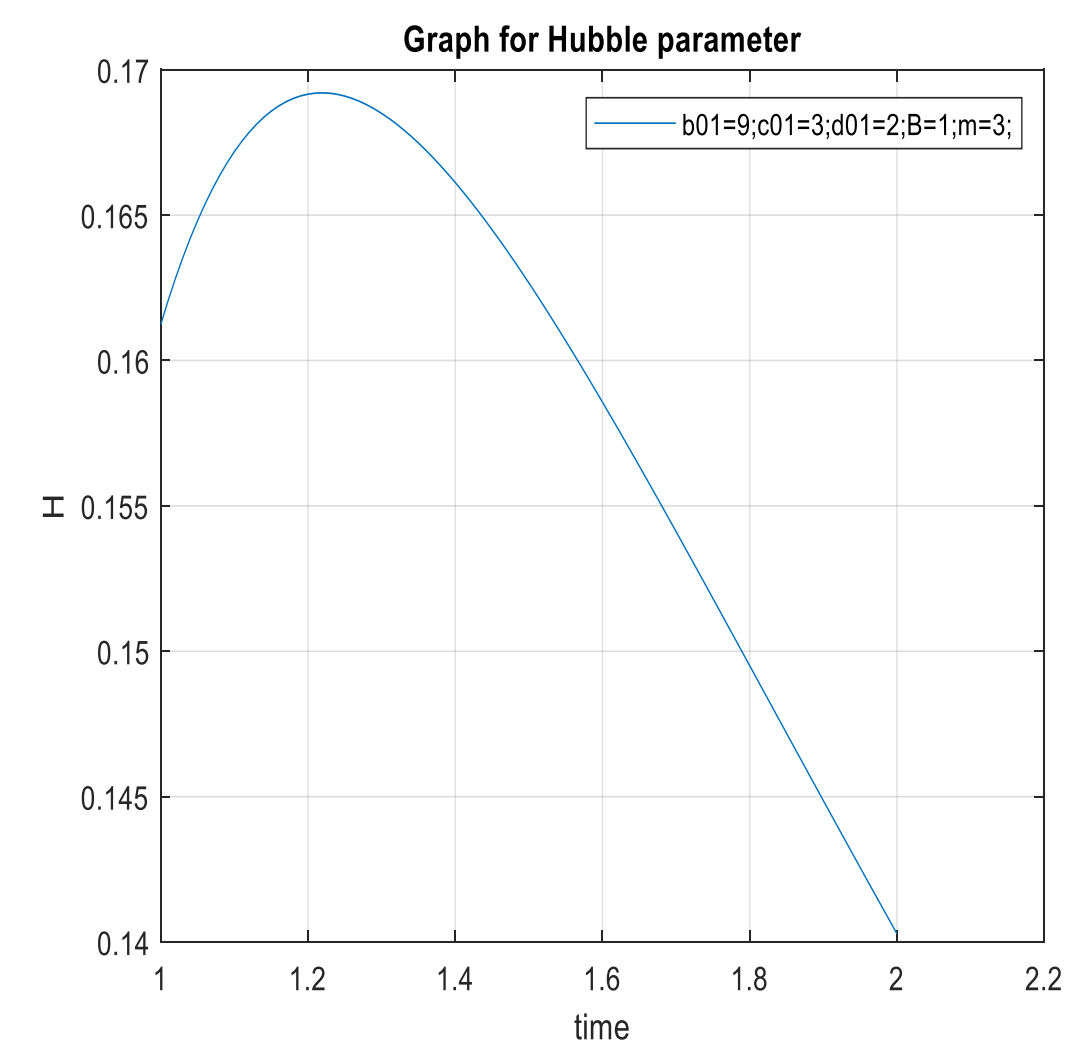

**Figure 23: Graph for Hubble parameter**

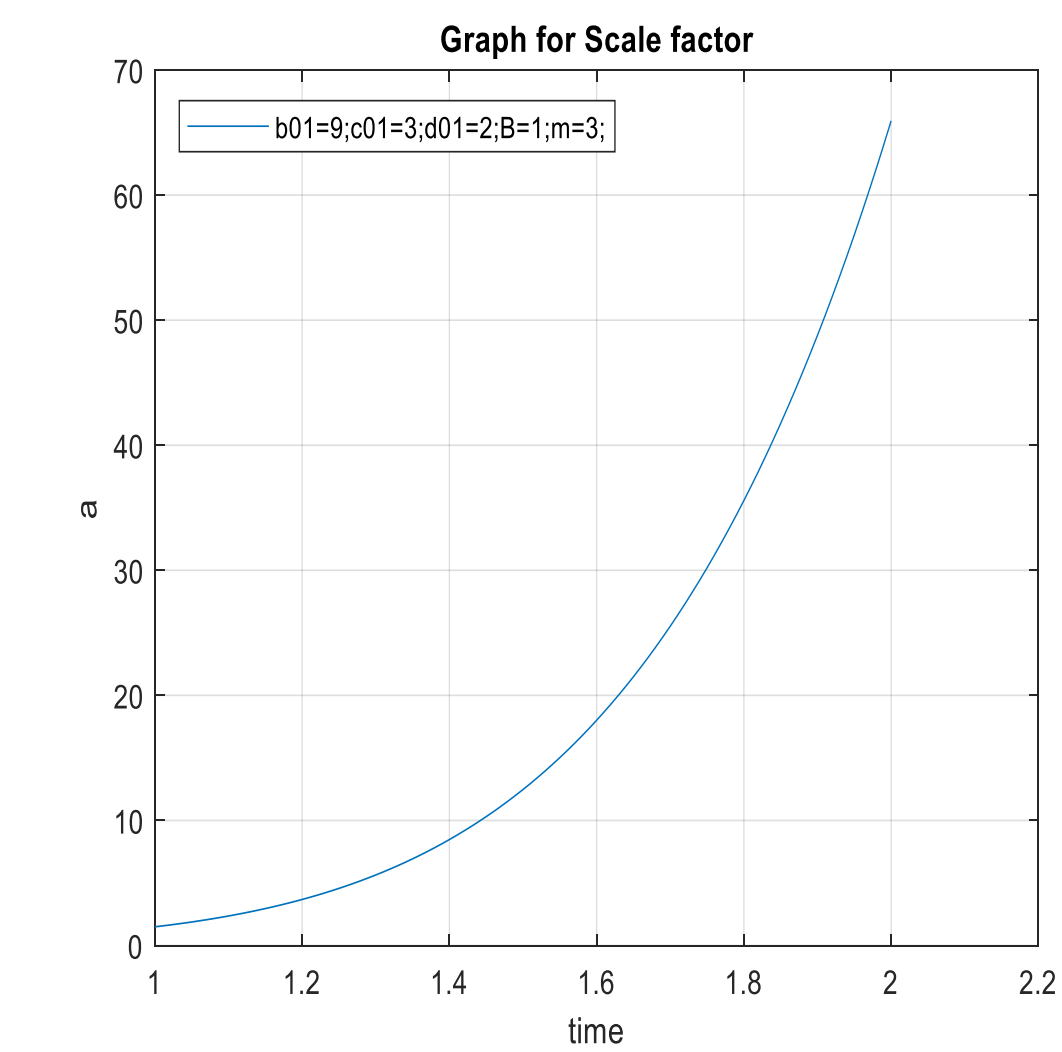

**Figure 24: Graph for reconstructed scale factor**

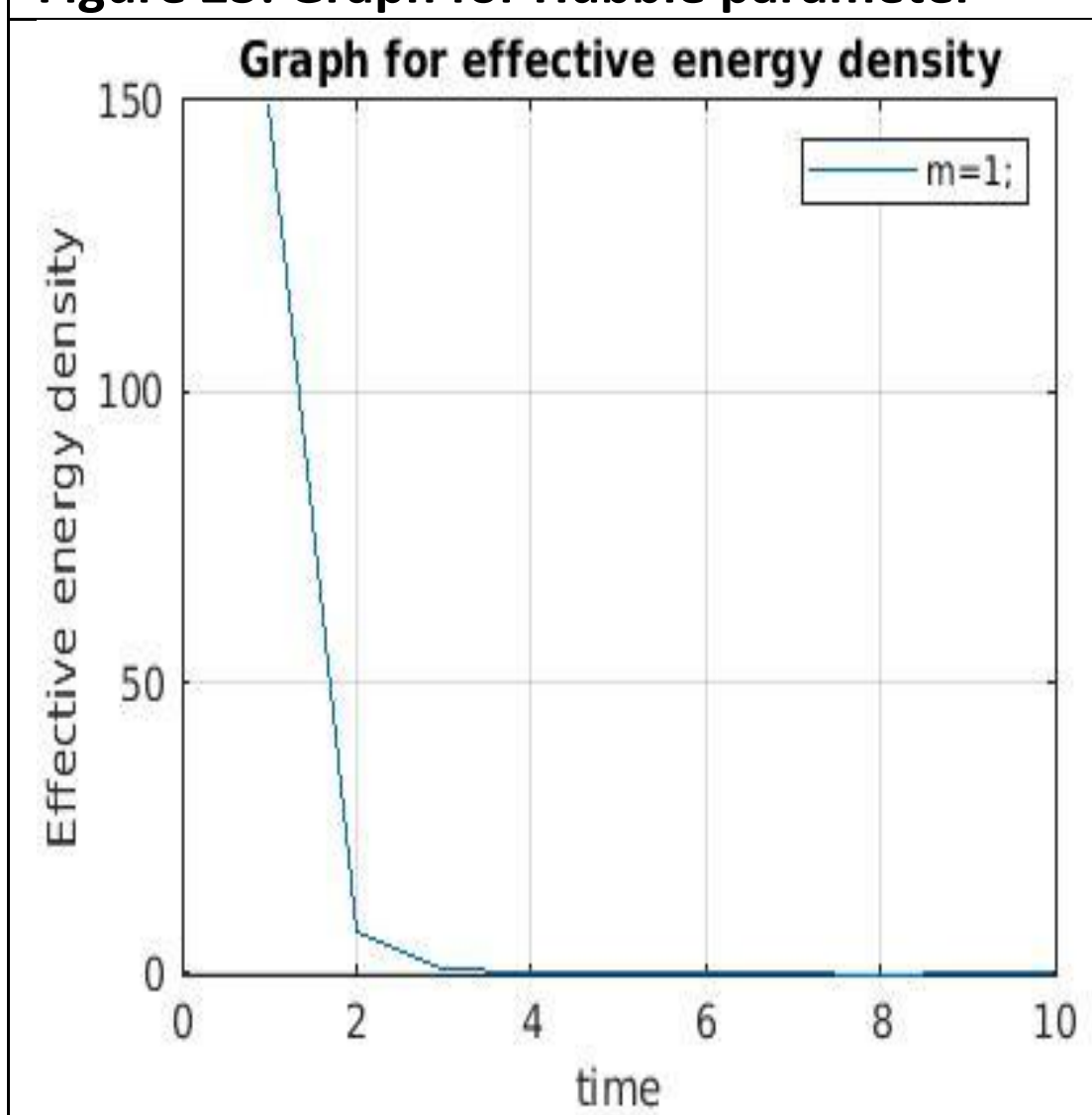

**Figure 25: Type I effective energy density**

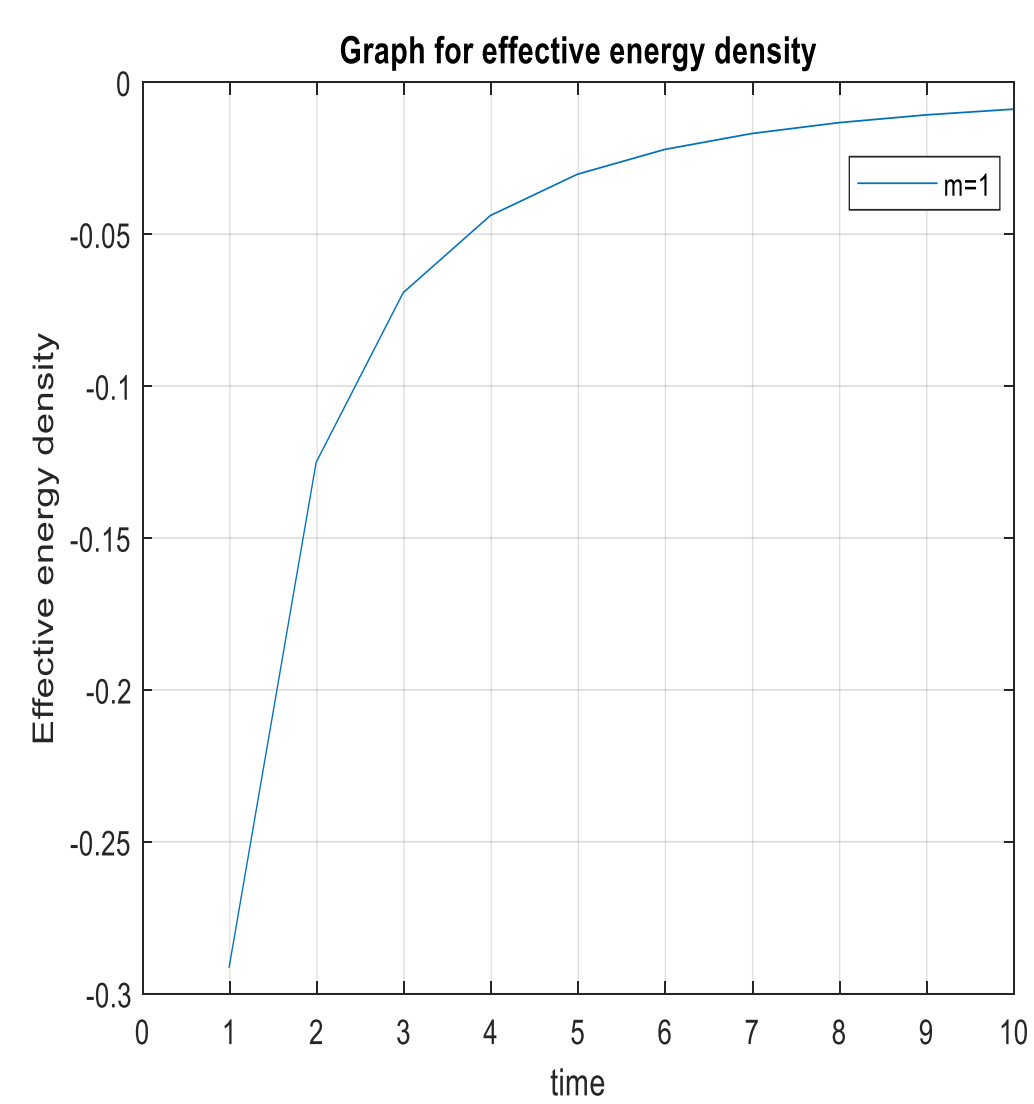

**Figure 26: Type II effective energy density**

**Here the finite time future singularity problems have been resolved with both scale factor and effective density. The Hubble parameter evolution follows the pseudo Rip evolution mechanism of universe. [110,111].**

**From the graph of scale factor, we can observe that for 1st type EOS we do not get any finite time future singularity w.r.t. the scale factor. From the graph of effective energy density $\rho_{eff}$ $vs$ $t$ we can say that the density will not have any finite time future singularity. So, from our model none of those 4 types singularity can be observed for any finite future time in phantom universe. (Here $\rho_{eff}$ $is\ equivalant\ to$ $\rho_{tot}$).**

**From the graphical representation of Hubble parameter, we may conclude that the value of this parameter will tends to zero when $t \to \infty$. If we consider $H = 0$ is an example of constant, then we may say that our model represents the pseudo rip evolution model of universe.**

**Type I EOS is stable in the accelerated expanding universe region and that is because only in this region we are getting positive scalar field energy density.**

**Type II EOS doesn't provide stable evolution of cosmic evolution and thus we did not get the positive effective energy density as well as positive scalar field energy density. Type II EOS also provides the resolution for finite time future singularity with respect to its effective density.**

**For both of the types the effective pressures are related with effective densities with their respective EOS and hence, their effective pressures types also provide the resolution of finite time future singularity problems.**

## 11. Thermodynamics of Inhomogeneous fluid for case I EOS

Here we'll be now discussing the thermodynamics of the type I case I Inhomogeneous fluid. Here We'll derive the temperature and entropy evolution with respect to volume scale factor for this type fluid. We'll also introduce thermodynamic stability idea to discuss the model stability and their conditions.[91-106]

### a. Internal energy

Here we have $p_d = a_2(a_1 + nt)^{-\alpha}\rho_d - c(a_1 + nt)^{-\beta}$ (49)

From the Hubble parameter-volume relation we may write

$$H^2 = \frac{m^2 n^2 {a_0}^{2/m}}{V^{2/3m}} = \frac{m^2 n^2}{(a_1+nt)^2} \Rightarrow (a_1 + nt) = \frac{V^{1/3m}}{{a_0}^{1/m}} \tag{50}$$

Again we have U = internal energy; so we get[92-104]

$$\rho_d = \frac{U}{V} \text{ and } p_d = -\left(\frac{\partial U}{\partial V}\right)_S \tag{51}$$

Here we are neglecting the time dependent coefficient of fluid density instead we use constant coefficient Inhomogeneous Fluid EOS. So $\alpha = 0$.

So we can write the EOS as follows;

$$\left(\frac{\partial U}{\partial V}\right)_S = -a_2\frac{U}{V} + c\,\frac{V^{-\beta/3m}}{a_0{}^{-\beta/m}} \tag{52}$$

So the internal energy is as follows;

$$U = \frac{3mca_0{}^{\beta/m}}{3m(a_1-1)-\beta}V^{-\left(\frac{3m+\beta}{3m}\right)} + constant * V^{-a_2} \tag{53}$$

So we can write further as;

$$\rho_d = \frac{3mca_0{}^{\beta/m}}{3m(a_1-1)-\beta}V^{-\left(\frac{6m+\beta}{3m}\right)} - a_2 * constant * V^{-(a_2+1)} \tag{54}$$

And

$$\mathrm{p}_d = a_2\left(\frac{3mca_0{}^{\frac{\beta}{m}}}{3m(a_1-1)-\beta}V^{-\left(\frac{6m+\beta}{3m}\right)} - a_2 * constant * V^{-(a_2+1)}\right) - c\,\frac{V^{-\beta/3m}}{a_0{}^{-\beta/m}} \tag{55}$$

So we get EOS parameter as follows;

$$w = \frac{p_d}{\rho_d} = \frac{a_2\left(\frac{3mca_0{}^{\frac{\beta}{m}}}{3m(a_1-1)-\beta}V^{-\left(\frac{6m+\beta}{3m}\right)} - a_2 * constant * V^{-(a_2+1)}\right) - c\frac{V^{-\beta/3m}}{a_0{}^{-\beta/m}}}{\frac{3mca_0{}^{\beta/m}}{3m(a_1-1)-\beta}V^{-\left(\frac{6m+\beta}{3m}\right)} - a_2 * constant * V^{-(a_2+1)}} \tag{56}$$

So from the above derivations we can find the volume variation of fluid density, pressure and EOS parameter.

### b. Temperature

Now from certain derivation we may write

$$\frac{\dot{T}}{T} = -3\frac{\dot{a}}{a}\frac{\partial p_d}{\partial \rho_d} \tag{57}$$

$$\text{Or, } \frac{dT}{T} = -\frac{dV}{V}\frac{\partial p_d}{\partial \rho_d} \tag{58}$$

Again we know that $\frac{\partial p_d}{\partial \rho_d} = a_2 + \frac{c\beta V^{-\left(\frac{\beta}{3m}+1\right)}}{3ma_0{}^{-\beta/m}}\frac{\partial V}{\partial \rho_d}$ (59)

Also from the energy conservation equation we get;

$$\dot{\rho}_d + 3H(\rho_d + p_d) = -3H\rho_d\delta \tag{60}$$

$$\text{Or, } \frac{\partial V}{\partial \rho_d} = -\frac{V}{(1+w+\delta)\rho_d} \tag{61}$$

So we may now write;

$$\frac{dT}{T} = -\frac{dV}{V}\left(a_2 - \frac{c\beta V^{-\left(\frac{\beta}{3m}\right)}}{3ma_0{}^{-\beta/m}(1+w+\delta)\rho_d}\right) \tag{62}$$

$$\text{Or, } T = \exp\left(-\int_0^V \frac{dV}{V}\left(a_2 - \frac{c\beta V^{-\left(\frac{\beta}{3m}\right)}}{3m{a_0}^{-\beta/m}(1+w+\delta)\rho_d}\right)\right) \tag{63}$$

This expression will provide us the scale factor dependent evolution of temperature which will be useful in CMBR data analysis.

### c. Entropy

Now from thermodynamic Maxwell relation we know that $T = \left(\frac{\partial U}{\partial S}\right)_V$

So we get from the definition of internal energy as follows;

$$T = V^{-a_2} * \left(\frac{\partial constant}{\partial S}\right)_V \tag{64}$$

From dimensional analysis we get;

$$[U] = [T]\,[S], \text{ and} [U] = [constant]\,[V]^{-a_2} \tag{65}$$

Now the constant is nothing but an integration constant which is a variable of entropy. So we get;

$$\text{Constant} = \tau v^{a_2} S \ \text{ and } \left(\frac{\partial constant}{\partial S}\right)_V = \tau v^{a_2} \tag{66}$$

So by further calculation we get;

$$S = \frac{constant}{T} V^{-a_2}$$

$$\text{Or; } S = \frac{constant}{\exp\left(-\int_0^V \frac{dV}{V}\left(a_2 - \frac{c\beta V^{-\left(\frac{\beta}{3m}\right)}}{3m{a_0}^{-\beta/m}(1+w+\delta)\rho_d}\right)\right)} V^{-a_2} \tag{67}$$

Thus we can derive the variable entropy for our cosmic fluid.

### d. Thermodynamics stability

Now let's discuss the thermodynamic stability of our model. To get thermodynamically stable model, the model need to satisfy the following rules.

Rule I: $\left(\frac{\partial p_d}{\partial V}\right)_S < 0$ (68)

Rule II: $\left(\frac{\partial p_d}{\partial V}\right)_T < 0$ (69)

Rule III: $c_v = T\left(\frac{\partial S}{\partial T}\right)_V > 0$ (70)

Now we see that,

$$\left(\frac{\partial p_d}{\partial V}\right)_S = a_2\left(-\frac{(6m+\beta)c{a_0}^{\frac{\beta}{m}}}{3m(a_1-1)-\beta} V^{-\left(\frac{6m+\beta}{3m}\right)} + a_2(a_2+1) * constant * V^{-(a_2+1)}\right) + \frac{c\beta}{3m}\frac{V^{-\left(\frac{\beta}{3m}-1\right)}}{{a_0}^{-\beta/m}} \tag{71}$$

So $\left(\frac{\partial p_d}{\partial V}\right)_S < 0$ for $constant \ll 0$, c > 0, and $a_2(a_2+1) > 0$ or, c $\ll$ 0 and $3m(a_1-1)-\beta \ll 0$

Again we get;

$$\mathrm{p}_d = a_2\left(\frac{3mca_0^{\frac{\beta}{m}}}{3m(a_1-1)-\beta}V^{-\left(\frac{6m+\beta}{3m}\right)} - a_2\frac{TS}{V}\right) - c\frac{V^{-\beta/3m}}{a_0^{-\beta/m}} \tag{72}$$

So $$\left(\frac{\partial p_d}{\partial V}\right)_T = a_2\left(-\frac{(6m+\beta)ca_0^{\frac{\beta}{m}}}{3m(a_1-1)-\beta}V^{-\left(\frac{6m+\beta}{3m}\right)} + a_2\frac{TS}{V^2} + a_2{}^2\frac{constant}{V^{a_2+1}}\right) + \frac{c\beta}{3m}\frac{V^{-\left(\frac{\beta}{3m}-1\right)}}{a_0^{-\beta/m}} \tag{73}$$

So $\left(\frac{\partial p_d}{\partial V}\right)_T < 0$ for $c < 0$, $a_2 \ll 0$ and $constant > 0$ or, $constant \ll 0$ and $a_2 > 0$, $c \ll 0$

And also

$$c_v = T\left(\frac{\partial S}{\partial T}\right)_V = -\frac{constant}{T}V^{-a_2} \tag{74}$$

So $c_v = T\left(\frac{\partial S}{\partial T}\right)_V > 0$ for constant < 0

Therefore all the stability rules are being satisfied in our calculation. The above derivations and conditions provides us one conclusion that shows negative entropy. In general entropy can't be negative for a running system. In this case the ideal fluid is interacting with scalar field where energy $3H\rho_d\delta$ is transferring from ideal fluid to scalar field. That's why the energy density in ideal fluid dominated universe is decreasing with increasing scalar field potential. That's why the entropy is becoming negative. Therefore, we may conclude that our model is thermodynamically stable only when the ideal fluid is interacting with some other field by some energy transition.

## 12. Thermodynamics of Inhomogeneous fluid for case II EOS

Here we'll be discussing the thermodynamics of the type I case II Inhomogeneous fluid. We'll derive the temperature and entropy evolution with respect to volume scale factor. We'll introduce thermodynamic stability idea to discuss the model stability and their conditions.[91-106]

### a. Internal energy

Here we have $\mathrm{p}_d = A\rho_d + BH^2$

From the Hubble parameter-volume relation we may write

$$H^2 = \frac{m^2n^2a_0{}^{2/m}}{V^{2/3m}}$$

Again we have U = internal energy; so we get

$$\rho_d = \frac{U}{V} \text{ and } \mathrm{p}_d = -\left(\frac{\partial U}{\partial V}\right)_S$$

So we can write the EOS as follows;

$$\left(\frac{\partial U}{\partial V}\right)_S = -A\frac{U}{V} - \frac{Bm^2n^2a_0{}^{2/m}}{V^{2/3m}} \tag{75}$$

So the internal energy is as follows;

$$U = -\frac{3Bm^2n^2a_0^{\frac{2}{m}}}{3m(1-A)-2}V^{\frac{3m-2}{3m}} + constant * V^{-A} \tag{76}$$

So we can write further as;

$$\rho_d = -\frac{3Bm^2n^2a_0^{\frac{2}{m}}}{3m(1-A)-2}V^{-\frac{2}{3m}} + constant * V^{-(A+1)} \tag{77}$$

And

$$p_d = Bm^2n^2a_0^{\frac{2}{m}}\left[1 - \frac{3A}{3m(1-A)-2}\right]V^{-\frac{2}{3m}} + A * constant * V^{-(A+1)} \tag{78}$$

So we get EOS parameter as follows;

$$w = \frac{p_d}{\rho_d} = \frac{Bm^2n^2a_0^{\frac{2}{m}}\left[1-\frac{3A}{3m(1-A)-2}\right]V^{-\frac{2}{3m}}+A*constant*V^{-(A+1)}}{-\frac{3Bm^2n^2a_0^{\frac{2}{m}}}{3m(1-A)-2}V^{-\frac{2}{3m}}+constant*V^{-(A+1)}} \tag{79}$$

So from the above derivations we can find the volume variation of fluid density, pressure and EOS parameter.

### b. Temperature

Now from certain derivation we may write[91]

$$\frac{\dot{T}}{T} = -3\frac{\dot{a}}{a}\frac{\partial p_d}{\partial \rho_d}$$

$$\text{Or, } \frac{dT}{T} = -\frac{dV}{V}\frac{\partial p_d}{\partial \rho_d}$$

Again we know that $\frac{\partial p_d}{\partial \rho_d} = A + 2BH\frac{\partial H}{\partial \rho_d}$ (80)

And $\frac{\partial H}{\partial \rho_d} = -\frac{mna_0{}^{1/m}}{3m}V^{-\frac{(1+3m)}{3m}}\frac{\partial V}{\partial \rho_d}$ (81)

Also from the energy conservation equation (60) and (61) we get;

$$\dot{\rho}_d + 3H(\rho_d + p_d) = -3H\rho_d\delta$$

$$\text{Or, } \frac{\partial V}{\partial \rho_d} = -\frac{V}{(1+w+\delta)\rho_d}$$

So we may now write;

$$\frac{dT}{T} = -\frac{dV}{V}\left(A + 2B\left(mna_0^{\frac{1}{m}}V^{-\frac{1}{3m}}\right)\left(-\frac{mna_0^{\frac{1}{m}}}{3m}V^{-\frac{(1+3m)}{3m}}\left(-\frac{V}{(1+w+\delta)\rho_d}\right)\right)\right) \tag{82}$$

Or,

$$\frac{dT}{T} = -\frac{dV}{V}\left(A + 2B\left(\frac{mn^2 a_0^{\frac{2}{m}}}{(1+w+\delta)\rho_d} V^{-\frac{2}{3m}}\right)\right) \quad (83)$$

$$\text{Or, } T = \exp\left(-\int_0^V \frac{dV}{V}\left(A + 2B\left(\frac{mn^2 a_0^{\frac{2}{m}}}{(1+w+\delta)\rho_d} V^{-\frac{2}{3m}}\right)\right)\right) \quad (84)$$

This expression will provide us the scale factor dependent evolution of temperature which will be useful in CMBR data analysis.

### c. Entropy

Now from thermodynamic Maxwell relation we know that $T = \left(\frac{\partial U}{\partial S}\right)_V$

So we get from the definition of internal energy as follows;

$$T = V^{-A} * \left(\frac{\partial constant}{\partial S}\right)_V$$

From dimensional analysis we get;

[U] = [T] [S], and [U] = [constant] $[V]^{-A}$

Now the constant is nothing but an integration constant which is a variable of entropy. So we get;

Constant = $\tau v^A S$ and $\left(\frac{\partial constant}{\partial S}\right)_V = \tau v^A$

So by further calculation we get;

$$S = \frac{constant}{T} V^{-A}$$

$$\text{Or; } S = \frac{constant}{\exp\left(-\int_0^V \frac{dV}{V}\left(A+2B\left(\frac{mn^2 a_0^{\frac{2}{m}}}{(1+w+\delta)\rho_d} V^{-\frac{2}{3m}}\right)\right)\right)} V^{-A} \quad (85)$$

Thus we can derive the variable entropy for our cosmic fluid.

### d. Thermodynamics stability

Now lets discuss the thermodynamic stability of our model. To get thermodynamically stable model, the model need to satisfy the following rules.

Rule I: $\left(\frac{\partial p_d}{\partial V}\right)_S < 0$

Rule II: $\left(\frac{\partial p_d}{\partial V}\right)_T < 0$

Rule III: $c_v = T\left(\frac{\partial S}{\partial T}\right)_V > 0$

Now we see that,

$$\left(\frac{\partial p_d}{\partial V}\right)_S = -\frac{2Bmn^2 a_0^{\frac{2}{m}}}{3}\left[1-\frac{3A}{3m(1-A)-2}\right]V^{-\frac{(2+3m)}{3m}} - (A+1)*constant*V^{-(A+2)} \quad (86)$$

So $\left(\frac{\partial p_d}{\partial V}\right)_S < 0$ for B > 0; (A+1) > 0 and $\left[1-\frac{3A}{3m(1-A)-2}\right] > 0$ (87)

Again we get;

$$p_d = Bm^2n^2a_0^{\frac{2}{m}}\left[1-\frac{3A}{3m(1-A)-2}\right]V^{-\frac{2}{3m}} + \frac{ATS}{V} \quad (88)$$

So
$$\left(\frac{\partial p_d}{\partial V}\right)_T = -\frac{2Bmn^2 a_0^{\frac{2}{m}}}{3}\left[1-\frac{3A}{3m(1-A)-2}\right]V^{-\frac{(2+3m)}{3m}} - \frac{ATS}{V^2} - \frac{A^2*constant}{V}V^{-(A+1)} \quad (89)$$

So $\left(\frac{\partial p_d}{\partial V}\right)_T < 0$ for B > 0; (A+1) > 0 and $\left[1-\frac{3A}{3m(1-A)-2}\right] > 0$ (90)

And also

$$c_v = T\left(\frac{\partial S}{\partial T}\right)_V = -\frac{constant}{V^A}\frac{1}{T} \quad (91)$$

So $c_v = T\left(\frac{\partial S}{\partial T}\right)_V > 0$ for constant < 0 (92)

Therefore all the stability rules are being satisfied in our calculation. The above derivations and conditions provides us one conclusion that shows negative entropy. In general entropy can't be negative for a running system. In this case the ideal fluid is interacting with scalar field where energy $3H\rho_d\delta$ is transferring from ideal fluid to scalar field. That's why the energy density in ideal fluid dominated universe is decreasing with increasing scalar field potential. That's why the entropy is becoming negative. Therefore we may conclude that our model is thermodynamically stable only when the ideal fluid is interacting with some other field by some energy transition.

## 13. Concluding Remarks:

We have discussed two different cases of a single type of inhomogeneous EOS and studied its different energy conditions. Even after satisfying all the energy conditions we get an accelerating universe from the calculations even in presence of attractive gravity. In order to satisfy the energy conditions we have got several values of constant that must be satisfied. For both the cases of EOS we have derived the time variation of gravitational constant, scalar field, and scalar field potential and cosmological constant. We have also introduced a new time dependent function of scale factor and Hubble parameter which resolves the Finite time future singularity for phantom era as well as initial time singularity. While discussing the thermodynamics of the two models we get expression for internal energy, temperature and entropy which are dependent on scale factor.

**The Expression of G (equation 38) in the paper has been computed by numerical integration since it is difficult to compute its actual function through normal integration. The present value of G is of the order $10^{-11}$, and the present age of the universe is considered to be 13.6 billion years. So to put some viable constraints on model parameters, we must work with such huge numbers in the programing, which was impossible for us. However, we can conclude from the**

**graphs that for type I EOS that the time variation of G provides the increasing nature where it should have almost constant nature w.r.t. time in early universe. This proves that the gravity geometry becoming more effective as time increases.**

**From Type II EOS we observe completely opposite nature as of type I. Here G provides decreasing nature with constant evolution on late time universe. Which says that the G must become almost constant in late time universe with lower value than the early universe. Although, this is stable only in decelerating phase of universe.**

**We can also observe from figure 8 that the scalar field energy density is decreasing with time. The late time increasing nature of G and the decreasing nature of scalar field density provides the proof that gravitational constant is inversely proportional to the evolution of scalar field energy density. This can provide a relation between scalar field and G that can reconstruct the scalar tensor theory of gravity. This increase of gravity G can be caused due to both the decrease of scalar field energy density and interaction between the multiple fluid that is discussed in earlier sections.**

**Reference**

1. Ren, J., Meng, X.H. and Zhao, L., 2007. Hamiltonian formalism in Friedmann cosmology and its quantization. *Physical Review D*, *76*(4), p.043521.
2. Verma, M.K. and Ram, S., 2011. Bianchi-Type VI 0 Bulk Viscous Fluid Models with Variable Gravitational and Cosmological Constants. *Applied Mathematics*, *2*(03), p.348..
3. Biswas, D., 2013. An Exact Scalar Field Inflationary Cosmological Model Which Solves Cosmological Constant Problem, Dark Matter Problem and Other Problems of Inflationary Cosmology..
4. Copeland, E.J., Sami, M. and Tsujikawa, S., 2006. Dynamics of dark energy. *International Journal of Modern Physics D*, *15*(11), pp.1753-1935.
5. Pimentel, O.M., Lora-Clavijo, F.D. and González, G.A., 2016. The energy-momentum tensor for a dissipative fluid in general relativity. *General Relativity and Gravitation*, *48*(10), p.124.
6. Harko, T., Lobo, F.S., Nojiri, S.I. and Odintsov, S.D., 2011. f (R, T) gravity. *Physical Review D*, *84*(2), p.024020.
7. Steinhardt, P.J., 2003. A quintessential introduction to dark energy. *Philosophical Transactions of the Royal Society of London. Series A: Mathematical, Physical and Engineering Sciences*, *361*(1812), pp.2497-2513.
8. Hughes, J., 2019. THE QUINTESSENTIAL DARK ENERGY THEORY: QUINTESSENCE.
9. Beesham, A., 1994. Bianchi type I cosmological models with variableG and Λ. *General relativity and gravitation*, *26*(2), pp.159-165.
10. Oscar M. Pimentel · F. D. Lora-Clavijo · Guillermo A. Gonz´alez;(2016);" The Energy-Momentum Tensor for a Dissipative Fluid in General Relativity"[arxiv:1606.01318v2]
11. Russell, E., Kılınç, C.B. and Pashaev, O.K., 2014. Bianchi I model: an alternative way to model the present-day Universe. *Monthly Notices of the Royal Astronomical Society*, *442*(3), pp.2331-2341.
12. Salih, M., 2009. A Canonical Quantization formalism of curvature squared action. *arXiv preprint arXiv:0901.2548*.

13. Lapchinskii, V.G. and Rubakov, V.A.E., 1977. Quantum gravitation: Quantization of the Friedmann model. *Theoretical and Mathematical Physics*, *33*(3), pp.1076-1084.
14. Faraoni, V. and Cooperstock, F.I., 2003. On the total energy of open Friedmann-Robertson-Walker universes. *The Astrophysical Journal*, *587*(2), p.483.
15. Elbaz, E., Novello, M., Salim, J.M., da Silva, M.M. and Klippert, R., 1997. Hamiltonian formulation of FRW equations of cosmology. *General Relativity and Gravitation*, *29*(4), pp.481-487.
16. Alvarenga, F.G. and Lemos, N.A., 1998. Dynamical vacuum in quantum cosmology. *General Relativity and Gravitation*, *30*(5), pp.681-694.
17. Monerat, G.A., Silva, E.V., Oliveira-Neto, G.D., Ferreira Filho, L.G. and Lemos, N.A., 2005. Notes on the quantization of FRW model in the presence of a cosmological constant and radiation. *Brazilian journal of physics*, *35*(4b), pp.1106-1109.
18. Salih, M., 2009. A Canonical Quantization formalism of curvature squared action. *arXiv preprint arXiv:0901.2548*.
19. Peres, A., 1999. Critique of the Wheeler-DeWitt equation. In *On Einstein's path* (pp. 367-379). Springer, New York, NY.
20. Alvarenga, F.G., Fabris, J.C., Lemos, N.A. and Monerat, G.A., 2002. Quantum cosmological perfect fluid models. *General Relativity and Gravitation*, *34*(5), pp.651-663.
21. Neves, C., Monerat, G.A., Corrêa Silva, E.V., Ferreira Filho, L.G. and Oliveira-Neto, G., 2011. Canonical transformation for stiff matter models in quantum cosmology. In *International Journal of Modern Physics: Conference Series* (Vol. 3, pp. 324-328). World Scientific Publishing Company.
22. Pedram, P., 2009. On the conformally coupled scalar field quantum cosmology. *Physics Letters B*, *671*(1), pp.1-6.
23. Lemos, N.A., 1996. Radiation-dominated quantum Friedmann models. *Journal of Mathematical Physics*, *37*(3), pp.1449-1460.
24. S. Sadhukhan, Quintessence model calculations for bulk viscous fluid and low value predictions of the coefficient of bulk viscosity. Int. J. Sci. Res. (IJSR) 9(3), 1419–1420 (2020). https://doi.org/10.21275/SR20327132301
25. A. Kar, S. Sadhukhan, Hamiltonian Formalism for Bianchi Type I Model for Perfect Fluid as Well as for the Fluid with Bulk and Shearing Viscosity. Basic and Applied Sciences into Next Frontiers (New Delhi Publishers, 2021) (ISBN: 978-81-948993-0-3)
26. A. Kar, S. Sadhukhan, Quintessence model with bulk viscosity and some predictions on the coefficient of bulk viscosity and gravita tional constant, recent advancement of mathematics in science and technology (2021) (ISBN: 978-81-950475-0-5)
27. Guth, A., 1982. Phys. Rev. D23: 347 (1981);
28. Kazanas, D., 1980. Dynamics of the universe and spontaneous symmetry breaking. *The Astrophysical Journal*, *241*, pp.L59-L63.
29. Sato, K., 1981. NORDITA-80-29, Jan 1980, published in Mon. *Not. Roy. Astron. Soc*, *195*, p.467..
30. Linde, A.D., 1982. A new inflationary universe scenario: a possible solution of the horizon, flatness, homogeneity, isotropy and primordial monopole problems. *Physics Letters B*, *108*(6), pp.389-393.
31. Albrecht, A., Steinhardt, P.J., Turner, M.S. and Wilczek, F., 1982. Reheating an inflationary universe. *Physical Review Letters*, *48*(20), p.1437.
32. Kamionkowski, M., 1998. New tests of inflation. *arXiv preprint astro-ph/9808004*.
33. Turner, M.S. and Widrow, L.M., 1988. Inflation-produced, large-scale magnetic fields. *Physical Review D*, *37*(10), p.2743.
34. Liddle, A.R., 1999. Observational tests of inflation. *arXiv preprint astro-ph/9910110*.

35. Lyth, D.H. and Stewart, E.D., 1992. The Curvature perturbation in power law (eg extended) inflation. *Physics Letters B*, *274*(2), pp.168-172.
36. Lucchin, F. and Matarrese, S., 1985. Power-law inflation. *Physical Review D*, *32*(6), p.1316.
37. Matravers, D., 2009. Steven Weinberg: Cosmology.
38. Spacetime and Geometry An introduction to General Relativity By Sean Carroll Addision-Wesley (2003)
39. Quantum Field theory in a nutshell (Second edition) By A. Zee, Princeton University Press (2010).
40. S. Perlmutter et al. Astrophysics J,517,565 (1999)
41. Filippenko, A.V. and Riess, A.G., 1998. Results from the high-z supernova search team. *Physics Reports*, *307*(1-4), pp.31-44.
42. Schmidt, B.P., Suntzeff, N.B., Phillips, M.M., Schommer, R.A., Clocchiatti, A., Kirshner, R.P., Garnavich, P., Challis, P., Leibundgut, B.R.U.N.O., Spyromilio, J. and Riess, A.G., 1998. The high-Z supernova search: measuring cosmic deceleration and global curvature of the universe using type Ia supernovae. *The Astrophysical Journal*, *507*(1), p.46.
43. General Relativity An Introduction for Physicist by M.P. Hobson. G.P. Efstathiou, And A.N. Lasenby. C.U.P. (2006).
44. Elementary Number Theory BY David M. Burton Tata Mcgraw-hill 6 th edition (2007).
45. Pi, S.Y., 1984. Inflation without tears: a realistic cosmological model. *Physical Review Letters*, *52*(19), p.1725.
46. Shafi, Q. and Vilenkin, A., 1984. Inflation with SU (5). *Physical Review Letters*, *52*(8), p.691.
47. Ryan, M.P. and Shepley, L.C., 2015. *Homogeneous relativistic cosmologies* (Vol. 65). Princeton University Press.
48. Barrow, J.D., 1984. Helium formation in cosmologies with anisotropic curvature. *Monthly Notices of the Royal Astronomical Society*, *211*(2), pp.221-227.
49. Ellis, G.F. and MacCallum, M.A., 1969. A class of homogeneous cosmological models. *Communications in Mathematical Physics*, *12*(2), pp.108-141.
50. Collins, C.B., 1971. More qualitative cosmology. *Communications in Mathematical Physics*, *23*(2), pp.137-158.
51. V. A. Ruban, "Preprint No. 412, Leningrade Institute of Nuclear Physics, B. P. Konstrantinova," Preprint, 1978.
52. Dunn, K.A. and Tupper, B.O.J., 1976. A class of Bianchi type VI cosmological models with electromagnetic field. *The Astrophysical Journal*, *204*, pp.322-329.
53. Lorenz, D., 1982. Tilted electromagnetic Bianchi type III cosmological solution. *Astrophysics and Space Science*, *85*(1-2), pp.59-61.
54. Roy, S.R. and Singh, J.P., 1983. Some Bianchi type VI 0 cosmological models with free gravitational field of the''magnetic''type. *Acta Physica Austriaca*, *55*(2), pp.57-66.
55. Ram, S., 1989. LRS Bianchi type I perfect fluid solutions generated from known solutions. *International journal of theoretical physics*, *28*(8), pp.917-921..
56. Ribeiro, M.B. and Sanyal, A.K., 1987. Bianchi VI0 viscous fluid cosmology with magnetic field. *Journal of mathematical physics*, *28*(3), pp.657-660.
57. Patel, L.K. and Koppar, S.S., 1991. Some Bianchi type VI 0 viscous fluid cosmological models. *The ANZIAM Journal*, *33*(1), pp.77-84..

58. Bali, R., Pradhan, A. and Amirhashchi, H., 2008. Bianchi type VI 0 magnetized barotropic bulk viscous fluid massive string universe in general relativity. *International Journal of Theoretical Physics*, *47*(10), pp.2594-2604.
59. Bali, R., Banerjee, R. and Banerjee, S.K., 2008. Bianchi type VI 0 magnetized bulk viscous massive string cosmological model in General Relativity. *Astrophysics and Space Science*, *317*(1-2), pp.21-26.
60. Bali, R., Banerjee, R. and Banerjee, S.K., 2009. Some LRS Bianchi Type VI 0 Cosmological Models with Special Free Gravitational Fields. *Electronic Journal of Theoretical Physics*, *6*(21).
61. L'DOVICH, Y.B.Z., 1962. The equation of state at ultrahigh densities and its relativistic limitations. *Soviet physics JETP*, *14*(5)..
62. Wagoner, R.V., 1970. Scalar-tensor theory and gravitational waves. *Physical Review D*, *1*(12), p.3209.
63. Linde, A.D., 1974. Is the Lee constant a cosmological constant. *JETP Lett*, *19*, p.183..
64. Vishwakarma, R.G., 2002. A Machian model of dark energy. *Classical and Quantum Gravity*, *19*(18), p.4747.
65. Kalligas, D., Wesson, P. and Everitt, C.W.F., 1992. Flat FRW models with variableG and Λ. *General Relativity and Gravitation*, *24*(4), pp.351-357.
66. Arbab, A.I., 1997. Cosmological models with variable cosmological and gravitational “Constants” and bulk viscous models. *General Relativity and Gravitation*, *29*(1), pp.61-74.
67. Vishwakarma, R.G., 1997. Some FRW models with variable G and. *Classical and Quantum Gravity*, *14*(4), p.945.
68. Pradhan, A. and Yadav, V.K., 2002. Bulk viscous anisotropic cosmological models with variable G and Λ. *International Journal of Modern Physics D*, *11*(06), pp.893-912.
69. Pradhan, A., Singh, A.K. and Otarod, S., 2006. FRW Universe with Variable $ G $ and $\Lambda $-Terms. *arXiv preprint gr-qc/0608107*.
70. Singh, J.P., Pradhan, A. and Singh, A.K., 2008. Bianchi type-I cosmological models with variable G and Λ-term in general relativity. *Astrophysics and Space Science*, *314*(1-3), pp.83-88.
71. Singh, C.P., Kumar, S. and Pradhan, A., 2006. Early viscous universe with variable gravitational and cosmological ‘constants’. *Classical and Quantum Gravity*, *24*(2), p.455..
72. Singh, J.P. and Tiwari, R.K., 2008. Perfect fluid Bianchi Type-I cosmological models with time varying G and Λ. *Pramana*, *70*(4), pp.565-574..
73. Singh, G.P. and Kotambkar, S., 2003. Higher-Dimensional Dissipative Cosmology with Varying G and Lambda. *Гравитация и космология*, *9*(3), pp.206-210..
74. Singh, G.P. and Kale, A.Y., 2009. Anisotropic bulk viscous cosmological models with variable G and Λ. *International Journal of Theoretical Physics*, *48*(4), pp.1177-1185..
75. Bali, R. and Tinker, S., 2009. Bianchi Type III Bulk Viscous Barotropic Fluid Cosmological Models with Variable G and Λ. *Chinese Physics Letters*, *26*(2), p.029802.
76. Verma, M.K. and Ram, S., 2010. Bulk viscous Bianchi type-III cosmological model with time-dependent G and Λ. *International Journal of Theoretical Physics*, *49*(4), pp.693-700.
77. Pradhan, A. and Bali, R., 2008. Magnetized Bianchi Type $ VI_ {0} $ Barotropic Massive String Universe with Decaying Vacuum Energy Density $\Lambda$. *arXiv preprint arXiv:0805.3469*..
78. Pradhan, A., Kumhar, S.S., Yadav, P. and Jotania, K., 2009. A New Class of LRS Bianchi Type ${\rm VI} _ {0} $ Universes with Free Gravitational Field and Decaying Vacuum Energy Density. *arXiv preprint arXiv:0907.4851*..

79. Maartens, R., 1995. Dissipative cosmology. *Classical and Quantum Gravity*, *12*(6), p.1455.
80. S. Weinberg, “In Gravitation and Cosmology,” Wiley, New York, 1972.
81. Murphy, G.L., 1973. Big-bang model without singularities. *Physical Review D*, *8*(12), p.4231.
82. Belinskii, V.A. and Khalatnikov, I.M., 1975. On the effect of viscosity on the nature of cosmological evolution. *ZhETF*, *69*, pp.401-413..
83. Ray, S., Mukhopadhyay, U. and Ghosh, P.P., 2007. Large number hypothesis: A review. *arXiv preprint arXiv:0705.1836*.
84. Weinberg, S. (1972). Gravitation and Cosmology, (Wiley, New York).
85. Abers, E. S., and Lee, B. W. (1973). Phys. Rep. 9, 1.
86. Langacker, P. (1981). Phys. Rep. 72, 185.
87. A. S. Al-Rawaf and M. O. Taha, Gen. Rel. Grav. 28 935 (1996).
88. Arbab I. Arbab, Class. Quantum Grav. 20 93 (2003); Astrophys. Space Sci. 291 141 (2004).
89. J. C. Carvalho, J. A. S. Lima and L. Waga, Phys. Rev. D 46 2404 (1992).
90. Kantha, L., 2016. A time-dependent and cosmological model consistent with cosmological constraints. *Advances in Astronomy*, *2016*.
91. Myrzakulov, R., Sebastiani, L. and Zerbini, S., 2013. Inhomogeneous Viscous Fluids in a Friedmann-Robertson-Walker (FRW) Universe. *Galaxies*, *1*(2), pp.83-95.
92. Jamil, M. and Rashid, M.A., 2008. Interacting dark energy with inhomogeneous equation of state. *The European Physical Journal C*, *56*(3), pp.429-434.
93. Khadekar, G.S., Raut, D. and Miskin, V.G., 2015. FRW viscous cosmology with inhomogeneous equation of state and future singularity. *Modern Physics Letters A*, *30*(29), p.1550144.
94. Nojiri, S.I. and Odintsov, S.D., 2005. Inhomogeneous equation of state of the universe: Phantom era, future singularity, and crossing the phantom barrier. *Physical Review D*, *72*(2), p.023003.
95. Štefančić, H., 2009. The solution of the cosmological constant problem from the inhomogeneous equation of state—a hint from modified gravity?. *Physics Letters B*, *670*(4-5), pp.246-253.
96. Brevik, I., Elizalde, E., Gorbunova, O. and Timoshkin, A.V., 2007. A FRW dark fluid with a non-linear inhomogeneous equation of state. *The European Physical Journal C*, *52*(1), pp.223-228.
97. Khadekar, G.S. and Raut, D., 2018. FRW viscous fluid cosmological model with time-dependent inhomogeneous equation of state. *International Journal of Geometric Methods in Modern Physics*, *15*(01), p.1830001.
98. Chakraborty, W. and Debnath, U., 2008. Interaction between scalar field and ideal fluid with inhomogeneous equation of state. *Physics Letters B*, *661*(1), pp.1-4.
99. Nojiri, S.I. and Odintsov, S.D., 2006. The new form of the equation of state for dark energy fluid and accelerating universe. *Physics Letters B*, *639*(3-4), pp.144-150.
100. Nojiri, S.I. and Odintsov, S.D., 2004. Final state and thermodynamics of a dark energy universe. *Physical Review D*, *70*(10), p.103522.
101. Nojiri, S.I., Odintsov, S.D. and Tsujikawa, S., 2005. Properties of singularities in the (phantom) dark energy universe. *Physical Review D*, *71*(6), p.063004.
102. Elizalde, E., Nojiri, S.I., Odintsov, S.D. and Wang, P., 2005. Dark energy: vacuum fluctuations, the effective phantom phase, and holography. *Physical Review D*, *71*(10), p.103504.
103. Chakraborty, S., Guha, S. and Panigrahi, D., 2019. Evolution of FRW universe in variable modified Chaplygin gas model. *arXiv preprint arXiv:1906.12185*.
104. Chakraborty, S. and Guha, S., 2019. Thermodynamics of FRW universe with Chaplygin gas models. *General Relativity and Gravitation*, *51*(11), pp.1-33.

105. Panigrahi, D. and Chatterjee, S., 2016. Thermodynamics of the variable modified Chaplygin gas. *Journal of Cosmology and Astroparticle Physics*, *2016*(05), p.052.
106. Panigrahi, D. and Chatterjee, S., 2017. Viability of variable generalised Chaplygin gas: a thermodynamical approach. *General Relativity and Gravitation*, *49*(3), p.35.
107. Moraes, P.H.R.S. and Sahoo, P.K., 2017. The simplest non-minimal matter–geometry coupling in the f (R, T) cosmology. *The European Physical Journal C*, *77*(7), pp.1-8.
108. Jamil, M. and Debnath, U., 2011. FRW cosmology with variable G and Λ. *International Journal of Theoretical Physics*, *50*(5), pp.1602-1613.
109. Barrow, J.D., 1993. Scalar-tensor cosmologies. *Physical Review D*, *47*(12), p.5329.
110. Brevik, I., Timoshkin, A.V. and Paul, T., 2021. The effect of thermal radiation on singularities in the dark universe.
111. arXiv preprint arXiv:2103.08430. Odintsov, S.D. and Oikonomou, V.K., 2017. Big bounce with finite-time singularity: The F (R) gravity description. International Journal of Modern Physics D, 26(08), p.1750085.